\documentclass[prc,superscriptaddress,noshowpacs,unsortedaddress,twocolumn,showpacs,preprintnumbers,amsmath,amssymb]{revtex4-2}

\usepackage[dvipdfmx]{graphicx}
\usepackage{amsmath,amssymb,times}
\usepackage{color}
\usepackage{ulem}
\usepackage{bm}
\usepackage{here}

\def\gsim{~\,\makebox(1,1){$\stackrel{>}{\widetilde{}}$}\,~}
\def\lsim{~\,\makebox(1,1){$\stackrel{<}{\widetilde{}}$}\,~}

\newcommand{\beq}{\begin{equation}}
\newcommand{\eeq}{\end{equation}}
\newcommand{\bea}{\begin{eqnarray}}
\newcommand{\eea}{\end{eqnarray}}

\newcommand{\bfi}[1]{\mbox{\boldmath $#1$}}

\newcommand{\vK}{{\bfi K}}

\newcommand{\vs}{{\bfi s}}

\newcommand{\vrr}{{\bfi r}}
\newcommand{\vR}{{\bfi R}}

\def\a{\alpha}
\def\b{\beta}

\begin{document}
\title{
Neutron skin thickness for $^{nat}$Pb,  $^{159}$Tb and effects of deformation on matter radii for  Mg, Na, Ar isotopes
}

\author{Shingo~Tagami}
\affiliation{Department of Physics, Kyushu University, Fukuoka 819-0395, Japan}

\author{Takaykui Myo}%
\affiliation{%
 General Education, Faculty of Engineering, Osaka Institute of Technology,
 Osaka 535-8585, Japan
}%

\author{Masanobu Yahiro}
\email[]{orion093g@gmail.com}
\affiliation{Department of Physics, Kyushu University, Fukuoka 819-0395, Japan}             

\date{\today}

\begin{abstract}
\begin{description}
\item[Background]
Data on reaction cross section $\sigma_{\rm R}$ of proton scattering are available for $^{nat}$Pb  (natural lead) and $^{159}$Tb. 
In addition, data on $\sigma_{\rm R}$ and/or interaction ones  
$\sigma_{\rm I}$ are available for  Mg, Na, Ar isotopes. 
\item[Purpose]
Our aim is to determine neutron skin thickness $r_{\rm skin}$
for $^{nat}$Pb,  $^{159}$Tb and to investigate effects of deformation for Mg, Na, Ar isotopes. 
\item[Methods]
 We use the chiral (Kyushu)  $g$-matrix folding model for lower energies and the folding model based on 
 the Love-Franey (LF) $t$-matrix for higher energies. 
\item[Results]
For $^{159}$Tb, our skin value is $r_{\rm skin}=0.273 \pm 0.073$~fm. 
As for $^{24}$Mg, our matter radius $r_{\rm m}(\sigma_{\rm R})=3.03	\pm 0.08$~fm determined from $\sigma_{\rm R}$  includes effects of deformation, whereas the corresponding $r_{\rm m}(\sigma_{\rm I})=2.79 \pm 0.15$~fm does not. 
Effects of deformation are seen for Mg isotopes,   $^{21, 23}$Na, 
$^{40}$Ar. 
\item[Conclusion]
Our result $r_{\rm skin}=0.271 \pm 0.008$~fm for $^{nat}$Pb agrees with $r_{\rm skin}^{208}({\rm PREX\text{-}II}) = 0.283\pm 0.071$~fm for  $^{208}$Pb, where we assume $A=208$ for $^{nat}$Pb. 
\end{description}
\end{abstract}

\maketitle


\section{Introduction}
\label{Introduction}

Horowitz {\it et al.} \cite{PRC.63.025501} proposed a direct measurement for neutron skin thickness $r_{\rm skin}$ = $r_{\rm n} - r_{\rm p}$, where 
$r_{\rm n}$ ($r_{\rm p}$) is the root-mean-square radii of neutrons 
(protons). 

The PREX collaboration reported~\cite{Adhikari:2021phr} 
\begin{equation}
r^{208Pb}_{skin}({\rm PREX\text{-}II}) 
= 0.283\pm 0.071\,{\rm fm}.
\label{Eq:Experimental constraint 208}
\end{equation}
The CREX group also did~\cite{CREX:2022kgg} 
\bea
&&r_{\rm skin}^{48Ca}({\rm CREX}) \nonumber \\
&&=0.121 \pm 0.026\ {\rm (exp)} \pm 0.024\ {\rm (model)}\,{\rm fm}. 
 \label{CREX-value}
 \eea

The values of   matter radii $r_{\rm m}({\rm exp})$, neutron radii $r_{\rm n}({\rm exp})$, 
$r_{\rm skin}({\rm exp})$, $r_{\rm p}({\rm exp})$ are tabulated in Table \ref{TW-prex} 
for $^{208}$Pb and $^{48}$Ca.

\begin{table}[htb]
\begin{center}
\caption
{Values of   matter radii $r_{\rm m}({\rm exp})$, neutron radii $r_{\rm n}({\rm exp})$, 
$r_{\rm skin}({\rm exp})$ 
together with $r_{\rm p}({\rm exp})$ 
deduced from the electron scattering~\cite{Brown:2013mga,Angeli:2013epw}. 
The radii are shown in units of fm.  
 }
\begin{tabular}{cccccc}
\hline\hline
 & . & $r_{\rm p}({\rm exp})$ & $r_{\rm m}({\rm exp})$ &  $r_{\rm n}{\rm exp})$ & $r_{\rm skin}({\rm exp})$ \\
\hline
  & PREX\text{-}II & $5.444$ & $5.617 \pm	0.044$ & $5.727 \pm 0.071$ & $0.283	\pm 0.071$ \\
  & CREX & $3.385$ & $3.456 \pm 0.030$ & $3.506 \pm 0.050$ & $0.121 \pm 0.050$ \\
\hline
\end{tabular}
 \label{TW-prex}
 \end{center} 
 \end{table}

In Ref.~\cite{Tagami:2020bee}, we extracted neutron skin thickness  $r_{\rm skin}^{208Pb}=0.278 \pm 0.035$~fm 
and  
matter radius $r_{\rm m}(\sigma_{\rm R})=5.614 \pm	0.022$~fm
 from  reaction cross sections $\sigma_{\rm R}$ on p+$^{208}$Pb scattering, using the chiral (Kyushu) $g$-matrix folding model~\cite{Toyokawa:2017pdd} with the proton and neutron densities calculated with Gogny-D1S HFB (D1S-GHFB) 
with angular momentum projection (AMP). 
Our skin value agrees with $r_{\rm skin}^{208Pb}({\rm PREX\text{-}II})$. 
The D1S-GHFB+AMP as a beyond mean field theory is very difficult for odd nuclei. The difficulty is shown In Ref.~\cite{PRC.101.014620}.

For a nucleus having mass number $A$, neutron number $N$, proton number $Z$,  
the $r_{\rm skin}({\rm exp})$ yields $r_{\rm n}({\rm exp})$,  $r_{\rm m}({\rm exp})$, when $r_{\rm p}({\rm exp})$ is obtainable from 
the electron scattering~\cite{Brown:2013mga,Angeli:2013epw}; note that $A~r_{\rm m}({\rm exp})^2=N r_{\rm n}({\rm exp})^2+Z r_{\rm p}({\rm exp})^2$. 
For the case of  $r_{\rm skin}({\rm exp})=0$, we obtain that 
$r_{\rm n}({\rm exp})=r_{\rm p}({\rm exp})=r_{\rm m}({\rm exp})$.
For the case of  $r_{\rm skin}({\rm exp}) \neq 0 $,  
the ratio $[r_{\rm n}({\rm exp})-r_{\rm m}({\rm exp})]/r_{\rm skin}({\rm exp}) $ are about 40~\% for both the $r^{208Pb}_{skin}({\rm PREX\text{-}II})$ and the $r_{\rm skin}^{48Ca}({\rm CREX})$.
In general, $r_{\rm n}({\rm exp})-r_{\rm m}({\rm exp})$ stems
from $r_{\rm skin}({\rm exp})$.

After this work, we extracted $r_{\rm skin}$ and matter radii 
$r_{\rm m}$ systematically.

\begin{figure}[H]
\begin{center}
 \includegraphics[width=0.4\textwidth,clip]{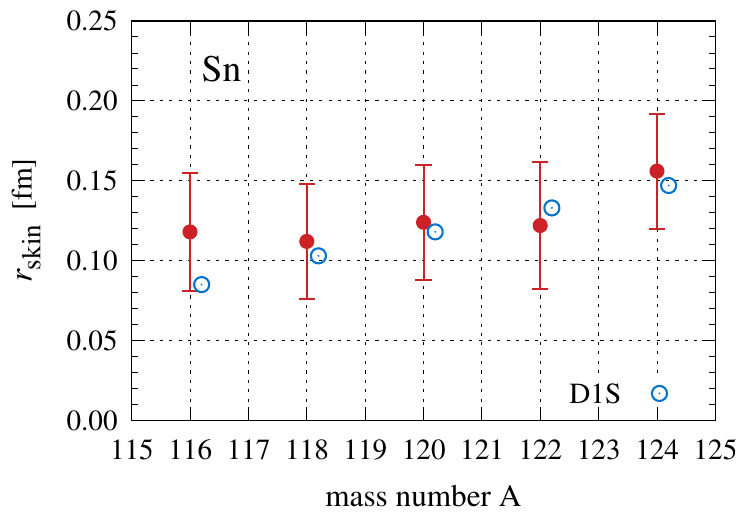}
 \caption{ 
 Skin values of $^{116,118,120,122,124}$Sn as a function of $A$. 
Closed circles with error-bar stand for our $r_{\rm skin}$ of p scattering for $^{116,118,120,122,124}$Sn in Ref.~\cite{TAGAMI2023106296}. 
Open circles denote the results calculated with 
D1S-GHFB+AMP.
 }
 \label{Fig-skins-compare}
\end{center}
\end{figure}

We extracted the $r_{\rm skin}$ of $^{116,118,120,122,124}$Sn from p+$^{116,118,120,122,124}$Sn scattering~\cite{TAGAMI2023106296}. 
As shown in Fig.~\ref{Fig-skins-compare},  the central values of the experimental data of Ref.~\cite{TAGAMI2023106296} slightly 
increase as $A$  changes from $A = 116$ to $A = 124$. Note that the experimental data do not show a clear dependence on $A$ 
 from $A = 118$ to $A = 124$, since the uncertainties are relatively large ($~30$\%) and the D1S-GHFB+AMP  points increase by $50$\% from $A = 118$ to $A = 124$.

The natural lead $^{nat}$Pb is composed of  $^{208}$Pb (54.4\%), 
$^{207}$Pb (22.1\%),
$^{206}$Pb (22.1\%),
$^{204}$Pb (1.4\%). 
Using our skin values for  $^{116,118,120,122,124}$Sn, we predict that the skin values  for $^{208,207,206,204}$Pb 
are almost constant as a function of $A$. 
Prediction A indicates that the skin value for  $^{nat}$Pb is close to $r_{\rm skin}^{208}({\rm PREX\text{-}II})$.

In this paper, we determine $r_{\rm skin}$ for $^{nat}$Pb  
from the data on $\sigma_{\rm R}$~\cite{Dietrich:2002swm,RENBERG197281,Cassels1954,Kirkby,
GOLOSKIE1962474,PhysRev.117.1334,GOODING1959241}.
Whether the  $r_{\rm skin}$ of $^{nat}$Pb is close to 
$r_{\rm skin}^{208Pb}({\rm PREX\text{-}II})$ or not? 
In addition, we determine $r_{\rm skin}$ for $^{159}$Tb  
from the data on $\sigma_{\rm R}$~\cite{ABEGG1979109} on p+$^{159}$Tb scattering,

Suzuki {\it et al.} measured interaction 
cross sections $\sigma_{\rm I}$ 
for $^{20\text{--}23, 25\text{--}32}$Na+ $^{12}$C  
and $^{20, 22, 23, 27, 29\text{--}32}$Mg+ $^{12}$C scattering at 950~MeV/u~\cite{Suzuki:1995yc,Suzuki:1998ixv} and determined matter radii $r_{\rm m}(\sigma_{\rm I})$~\cite{Suzuki:1998ixv}.  
The measurement for Na isotopes was the first direct comparison between neutron radii $r_{\rm n}$ and proton ones $r_{\rm p}$ over a wide range of neutron number $N$, where the charge radii were determined by isotope-shift measurements~\cite{Huber:1978zz}. 
As for $^{21, 23}$Na+$^{12}$C scattering, 
the reaction cross sections $\sigma_{\rm R}$ were measured by Zhang {\it et al.} at around 30~MeV/u~\cite{ZHANG2002303}.

There is an accumulation paper~\cite{Ozawa:2001hb} 
for $^{4}$He to $^{32}$Mg. 
Data on $\sigma_{\rm I}$ at high energies, 
those on $\sigma_{\rm R}$ at low energies  
and the extracted  $r_{\rm m}(\sigma_{\rm I})$ 
are accumulated; see Table 1 
for the $\sigma_{\rm I}$, Table 2 for the $\sigma_{\rm R}$, 
Table 3  for 
the $r_{\rm m}(\sigma_{\rm I})$ that were by using Glauber model in the optical limit.

After the accumulation paper, 
Ozawa {\it et. al.} measured $\sigma_{\rm I}$
for $^{32 \text{--} 40}$Ar+$^{12}$C scattering  
at around 950 MeV/u, and extracted 
the proton skin thickness $r_{\rm skin,p} = r_{\rm p}-r_{\rm n}$ 
and the  $r_{\rm m}(\sigma_{\rm I})$~\cite{Ozawa:2002av}, using 
the Glauber-model calculations in the optical-limit approximation.

As an observable of determining  $r_{\rm m}$, $r_{\rm skin}$, 
we can consider $\sigma_{\rm R}$, $\sigma_{\rm I}$. 
The $\sigma_{\rm R}({\rm exp})$  is related to 
$\sigma_{\rm I}({\rm exp})$ as 
\bea
\sigma_{\rm I}({\rm exp})=\sigma_{\rm R}({\rm exp})-\sigma_{\rm inela}
\label{Eq:inela}
\eea
for the total inelastic cross sections $\sigma_{\rm inela}$. 
The matter radius $r_{\rm m}(\sigma_{\rm I})$ extracted from $\sigma_{\rm I}$
is smaller than $r_{\rm m}(\sigma_{\rm R})$. There is possibility that 
$r_{\rm m}(\sigma_{\rm I})$ do not include effects of deformation, since large deformation induces strong excitation.

Takechi {\it et al.} made precise measurements 
on $\sigma_{\rm R}$ for $^{24\text{--}38}$Mg on a $^{12}$ C target 
at energies around 240 MeV/u~\cite{PhysRevC.90.061305}.  
Watanabe {\it et al.} determined $r_{\rm m}(\sigma_{\rm R})$ for $^{24\text{--}38}$Mg~\cite{Watanabe:2014zea}. 
The ground-state properties of Mg isotopes based on D1S-AMD (antisymmetrized molecular dynamics) are shown
in Ref.~\cite{Watanabe:2014zea}. 
For Mg isotopes, the $r_{\rm m}(\sigma_{\rm R})$ of Ref.~\cite{Watanabe:2014zea} and the $r_{\rm m}(\sigma_{\rm I})$ of Ref.~\cite{Ozawa:2001hb} were tabulated;   
Table~\ref{reference values-0} shows theses values for $^{24}$Mg.
There is non-negligible difference between them.  This point is investigated in this paper for Mg isotopes. 

\begin{table}[H]
\begin{center}
\caption
{Comparison between our values $r_{\rm m}(\sigma_{\rm R})$ 
of  Refs.~\cite{TAGAMI2023106675,Watanabe:2014zea,WAKASA2022105329} 
and $r_{\rm m}(\sigma_{\rm I})$ of  Ref.~\cite{Ozawa:2001hb}.  
The radii are shown in units of fm.  
 }
\begin{tabular}{cccccc}
\hline\hline
 &  &  & 
 $r_{\rm m}(\sigma_{\rm I})$ & $r_{\rm m}(\sigma_{\rm R})$ & \\
\hline
 $^{6}$He &  & & $2.48 \pm 0.03$ & $2.48 \pm  0.03$ & \\
 $^{8}$He &  & & $2.52 \pm  0.03$ & $2.53 \pm  0.02$ & \\
$^{12}$C & & & $2.35 \pm 0.02$ &   $2.352 \pm 0.013$ & \\
$^{24}$Mg & & & $2.79	\pm 0.15$ &   $3.03	\pm 0.08$ & \\
$^{27}$Al & & & $$ &   $2.936 \pm 0.012$ & \\
\hline
\end{tabular}
 \label{reference values-0}
 \end{center} 
 \end{table}

Takechi {\it et al.} made precise measurements on 
$\sigma_{\rm R}$ for $^{12}$C on $^{9}$Be, $^{12}$C, $^{27}$Al targets 
in the wide incident-energy range of $E_{\rm lab}=30\text{--}400$~MeV/u~\cite{PhysRevC.79.061601}. 
Using the precise data for  $^{12}$C+$^{12}$C scattering, 
we tested the Kyushu $g$-matrix folding model for  $^{12}$C+$^{12}$C scattering, and found that the model is reliable for 
$\sigma_{\rm R}$
in $30  \lsim E_{\rm lab} \lsim 100 $~MeV/u and $250  \lsim E_{\rm lab} \lsim 400 $~MeV/u~\cite{PRC.101.014620}. Note that no data is available in $ E_{\rm lab} \leq 30$~MeV/u.

The chiral nucleon-nucleon interactions used in the Kyushu $g$-matrix folding model have a cutoff of $\Lambda=550$~MeV; the cutoff effects appear $450 \lsim E_{\rm lab} \lsim 550$~MeV for the $\sigma_{\rm R}$ of 
$^{12}$C+$^{12}$C scattering.

We tested the Kyushu $g$-matrix folding model~\cite{Toyokawa:2017pdd} 
for proton scattering and found that 
the chiral (Kyushu) $g$-matrix folding model is reliable for 
$\sigma_{\rm R}({\rm exp})$ 
in $20  \lsim E_{\rm lab} \lsim 180$ MeV~\cite{Wakasa:2022ite}.
For this reason, the model is applicable for $E_{\rm lab} \lsim 450$~MeV.
 In  $ E_{\rm lab} \gsim 550$~MeV/u, we use the folding model based on  Love-Franey (LF) $t$-matrix~\cite{LF}.

Takechi {\it et al.} also measured $\sigma_{I}$ 
for $^{20-32}$Ne+$^{12}$C scattering at around 
240 MeV/u~\cite{TAKECHI2012357}. 
Minomo  {\it et. al.} used the D1S-AMD+RGM method to describe the tail of the last neutron of $^{31}$Ne, 
where RGM stands for the resonating group method and 
AMD is antisymmetrized molecular dynamics. The theoretical prediction reproduces the measured 
$\sigma_{I}$ of $^{28\mbox{--}32}$Ne with no adjustable parameter~\cite{PhysRevLett.108.052503}. 
The theoretical result indicates that the ground state of $^{31}$Ne is a  ``deformed'' halo nucleus with ${(3/2)}^{-}$.
The ground-state properties of Ne isotopes based on AMD are shown in Ref \cite{PhysRevC.85.064613}.

From the precise data on the $\sigma_{\rm R}$ of Ref.~\cite{PhysRevC.79.061601}, we extracted 
$r_{\rm m}(\sigma_{\rm R})$ for $^{12}$C and $^{27}$Al~\cite{TAGAMI2023106675}, using the chiral (Kyushu) $g$-matrix folding model~\cite{Toyokawa:2017pdd}. 
In Ref.~\cite{WAKASA2022105329}, 
we also determined $r_{\rm m}(\sigma_{\rm R})$ for $^{6,8}$He
from $\sigma_{\rm R}$ of p+$^{6,8}$He scattering at 700~MeV, 
using the folding model based on  Love-Franey (LF) $t$-matrix~\cite{LF}.
As shown in Table~\ref{reference values-0}, our values  for $^{6,8}$He are consistent with the corresponding values 
$r_{\rm m}(\sigma_{\rm I})$ shown 
in the accumulation paper.  
This consistency indicates that 
the Kyushu $g$-matrix folding model and the LF folding model are reliable.

In Ref.~\cite{Wakasa:2022ite}, we extracted  
$r_{\rm skin}$ and $r_{\rm m}(\sigma_{\rm R})$ from 
$\sigma_{\rm R}$ of p scattering on $^{208}$Pb, $^{58}$Ni, $^{48,40}$Ca, $^{12}$C targets, using the Kyushu $g$-matrix folding model.
As a way of determining a fine-tuning factor $f$, we proposed 
the ESP-f (experimental scaling procedure). The  ESP-f is a reliable way of $r_{\rm skin}$ and $r_{\rm m}(\sigma_{\rm R})$. 
Our skin values of  $^{208}$Pb and $^{48}$Ca 
are consistent with the PREX\text{-}II value~\cite{Adhikari:2021phr} and 
the CREX one~\cite{CREX:2022kgg}, respectively.

Tanaka {\it et al.} measured   
$\sigma_{\rm I}$ for $^{42-51}$Ca+ $^{12}$C scattering 
at 280~MeV/u, and 
determined  matter radii $r_{\rm m}(\sigma_{\rm I})$, neutron radii $r_{\rm n}(\sigma_{\rm I})$, neutron  skin thickness $r_{\rm skin}$ 
for $^{42-51}$Ca 
from the $\sigma_{\rm I}$, 
using the optical limit of the Glauber model with the Woods-Saxon densities~\cite{Tanaka:2019pdo}. 
We reanalyze the scattering to determine 
$r_{\rm m}(\sigma_{\rm I})$, 
$r_{\rm n}(\sigma_{\rm I})$, 
$r_{\rm skin}$ for $^{42-51}$Ca
in Ref.~\cite{TAKECHI2021104923}, 
using the Kyushu $g$-matrix folding model with the proton and 
neutron densities calculated 
with D1S-GHFB with and 
without AMP.  Our values are consistent with theirs.

In this paper, our aim is to determine neutron skin thickness $r_{\rm skin}$
for $^{nat}$Pb,  $^{159}$Tb and to investigate effects of deformation for Mg, Na, Ar isotopes. Our analyses are summarized blow.

\begin{description}

      \item[1)~p+$^{nat}$Pb scattering]
We extract  $r_{\rm skin}$ and $r_{\rm m}(\sigma_{\rm R})$ from  data $\sigma_{\rm R}$~\cite{Dietrich:2002swm,RENBERG197281,Cassels1954,Kirkby,
GOLOSKIE1962474,PhysRev.117.1334,GOODING1959241}  in $E_{\rm lab} =21.1 \sim 341$~MeV, using the Kyushu $g$-matrix folding model with the  D1S-GHFB+AMP proton and neutron densities. We assume $A=208$ for 
$^{nat}$Pb. 

      \item[2)~p+$^{159}$Tb scattering]
We extract  $r_{\rm skin}$ and $r_{\rm m}(\sigma_{\rm R})$ from  data $\sigma_{\rm R}$~\cite{ABEGG1979109}  in $20  \leq E_{\rm lab} \leq 47.5$~MeV, using the Kyushu $g$-matrix folding model with the  
D1S-GHFB densities.

     \item[3-1)~ ${^{20{\text--}32}}$Na+$^{12}$C scattering] 
    We determine $r_{\rm m}(\sigma_{\rm I})$ for ${^{20{\text--}32}}$Na on the data $\sigma_{\rm I}$~\cite{Suzuki:1995yc,Suzuki:1998ixv} at 950~MeV/u, using the LF $t$-matrix folding model with the Sly7 densities, where SLy7~\cite{Chabanat:1997un,Schunck:2016uvm,Matsuzaki:2021hdm} is an improved version of SLy4. 

         \item[3-2)~${^{21,23}}$Na+$^{12}$C scattering] 
    We determine $r_{\rm m}(\sigma_{\rm R})$ for $^{21, 23}$Na on data 
    $\sigma_{\rm R}$~\cite{ZHANG2002303} at around 30~MeV/u, using the Kyushu $g$-matrix folding model with the Sly7 densities.

          \item[4)~Comparison between $r_{\rm m}(\sigma_{\rm R})$ and 
          $r_{\rm m}(\sigma_{\rm I})$ for Mg isotopes] 
We analyzed  $^{24\text{-}38}$Mg+$^{12}$C scattering at 240~MeV/u and determined the 
 $r_{\rm m}(\sigma_{\rm R})$ for $^{24\text{-}38}$Mg~\cite{Watanabe:2014zea}.
Meanwhile,  Suzuki {\it et al.} measured $\sigma_{\rm I}$ 
for $^{20, 22, 23, 27, 29\text{--}32}$Mg+ $^{12}$C 
scattering at 950~MeV/u~\cite{Suzuki:1995yc,Suzuki:1998ixv} and determined matter radii $r_{\rm m}(\sigma_{\rm I})$~\cite{Suzuki:1998ixv}.  
We find that the $r_{\rm m}(\sigma_{\rm I})$ of Ref.~\cite{Suzuki:1998ixv} yield do not include effects of deformation, but  the $r_{\rm m}(\sigma_{\rm R})$ of Ref.~\cite{Watanabe:2014zea} include effects of deformation; 
see Fig.~\ref{Fig-Rm-Mg}. 
This is investigated with D1S-GHFB+AMP and  the D1S-GHFB+AMP in the spherical limit. 

       \item[5-1)~$^{32\text{--}40}$Ar+$^{12}$C scattering]
We determine $r_{\rm m}(\sigma_{\rm I})$ and $r_{\rm skin,p}$ for $^{32{\text--}40}$Ar  from the $\sigma_{\rm I}$
at around 950 MeV/u~\cite{Ozawa:2002av}, using the LF folding model with the Sly7 densities.

      \item[5-2)~p+$^{40}$Ar scattering]
We extract  $r_{\rm skin}$ and $r_{\rm m}(\sigma_{\rm R})$ from  data $\sigma_{\rm R}$~\cite{CARLSON198557}  in $22.9  \leq E_{\rm lab} \leq 46.9$~MeV, using the Kyushu $g$-matrix folding model with the  Sly7 densities.

\end{description}

Our model is formulated in Sec.~\ref{Sec-Framework}, 
and our results are shown in Sec.~\ref{Results}.  
Section \ref{Summar} is devoted to a summary.

\section{Model}
\label{Sec-Framework}

\subsection{The $g$-matrix folding model}

The $g$-matrix folding model~\cite{Brieva-Rook,Amos,CEG07, Sumi:2012fr, PhysRevC.88.054602,Egashira:2014zda,Watanabe:2014zea,Toyokawa:2014yma,Toyokawa:2015zxa,Toyokawa:2017pdd} is a standard way of 
determining $r_{\rm skin}$ and $r_{\rm m}$ from $\sigma_{\rm R}$ and
$\sigma_{\rm I}$. The formulation of the $g$-matrix folding model is established, as shown in Refs.~\cite{Brieva-Rook,Amos,CEG07, Sumi:2012fr, Egashira:2014zda,Watanabe:2014zea,Toyokawa:2014yma,Toyokawa:2015zxa,Toyokawa:2017pdd}.
In the model, the  potential is obtained by folding  the $g$-matrix with projectile (P) densities and target (T) ones.

The $g$-matrix  folding model consists of the single-folding model (SFM)  for 
nucleon-nucleus  scattering and the double-folding model (DFM) for nucleus-nucleus  scattering.
Satchler succeeded in reproducing most of elastic nucleus-nucleus  scattering with the 
the central direct and the central particle-exchange folding potential~\cite{Satchler:1979ni}. 
After the work,  the SFM has the central and the spin-orbit folding potential, whereas the DFM has   the central potential only.

Applying the folding model based on 
the Melbourne $g$-matrix~\cite{Amos}   for  
$\sigma_{\rm R}$ of Mg isotopes, we 
deduced the $r_{\rm m}(\sigma_{\rm R})$ for   
Mg isotopes~\cite{Watanabe:2014zea}, and discovered 
that $^{31}$Ne is a halo nucleus with large deformation~\cite{PhysRevLett.108.052503}. 

\begin{figure}[tbp]
\begin{center}
 \includegraphics[width=0.54\textwidth,clip]{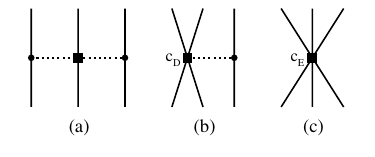}
 \caption{3NFs in NNLO. 
Diagram (a) corresponds 
to the Fujita-Miyazawa 2$\pi$-exchange 3NF \cite{PTP.17.360}, 
and diagrams (b) and (c) correspond to 1$\pi$-exchange and contact 3NFs.
The solid and dashed lines denote nucleon and pion propagations, 
respectively, and filled circles and squares stand for vertices. 
The strength of the filled-square vertex is often called $c_{D}$ 
in diagram (b) and $c_{E}$ in diagram (c). 
}
 \label{fig:diagram}
\end{center}
\end{figure}

Kohno calculated the $g$ matrix  for the symmetric nuclear matter, 
using the Brueckner-Hartree-Fock method with chiral N$^{3}$LO 2NFs and NNLO 3NFs~\cite{Koh13}. 
The chiral interactions mentioned above are explained in 
Fig.~\ref{fig:diagram}.
He set $c_D=-2.5$ and $c_E=0.25$ so that  the energy per nucleon can  become minimum 
at $\rho = \rho_{0}$~\cite{Toyokawa:2017pdd}. 

Toyokawa {\it et al.} localized the non-local chiral  $g$ matrix 
into three-range Gaussian forms by using the localization method proposed 
by the Melbourne group~\cite{von-Geramb-1991,Amos-1994,Amos}. 
The resulting local  $g$ matrix is called  ``Kyushu  $g$-matrix''. 

The  Kyushu $g$-matrix folding model is successful in reproducing $d\sigma/d\Omega$ and $A_y$
for polarized  proton scattering  on various targets at $E_{\rm lab}=65$~MeV~\cite{Toyokawa:2014yma}
and $d\sigma/d\Omega$ for $^4$He scattering at $E_{\rm lab}=72$~MeV/u~\cite{Toyokawa:2015zxa}. 
This is true for $\sigma_{\rm R}$ of $^4$He scattering 
in $E_{\rm lab}=30 \sim 200$~MeV/u~\cite{Toyokawa:2017pdd}.

We use the Kyushu $g$-matrix  folding model~\cite{Toyokawa:2017pdd} 
and the LF folding model for nucleus-nucleus (PT) scattering. 
nucleon-nucleus (NT) scattering

\subsection{Single-folding model}

We consider proton as a projectile. 
Proton elastic scattering can be 
described by the one-body Schr\"odinger equation,
\bea
(T_R + U ({\bm R})-E){\Psi}^{(+)}=0 , 
\label{schrodinger}
\eea
with a folded potential $U({\bm R})$, 
where 
$E$ is an  energy of incident proton in the center of mass system and 
$T_R$ is a kinetic-energy operator concerning the relative coordinate 
$\vR$ between an incident proton and a target. 
The optical potential $U$ can be divided into 
the central (CE), the spin-orbit (LS), and the Coulomb (Coul) component 
as follows:
\bea
U({\bm R})=U_{\rm CE}+U_{\rm LS} {\bfi L} \cdot {\bfi \sigma} 
+ V_{\rm Coul} . 
\eea

In the $g$-matrix folding model, $U$ consists of 
the direct and particle-exchange parts. The exchange part 
is nonlocal, but it can be localized with 
the Brieva-Rook approximation \cite{Brieva-Rook}. 
Validity of the approximation is shown in Ref. \cite{Minomo:2009ds}. 

The central part $U_{\rm CE}$ is then described by the sum of 
the direct component $U_{\rm CE}^{\rm DR}$ 
and the localized exchange component 
$U_{\rm CE}^{\rm EX}$~\cite{Brieva-Rook,CEG,CEG07},
\bea
U_{\rm CE} \equiv V_{\rm CE}+iW_{\rm CE}
=U_{\rm CE}^{\rm DR}+U_{\rm CE}^{\rm EX}
\eea
with 
\bea
\label{eq:UD}
U_{\rm CE}^{\rm DR}(\vR) \hspace*{-0.15cm} &=& \hspace*{-0.15cm} 
\!\! \sum_{\alpha=p,n} \! \int \!\! \rho_{\alpha}(\vrr) 
            g^{\rm DR}_{p\alpha}(s;\rho_{\alpha}) d \vrr, 
\label{U-DR}     
            \\
U_{\rm CE}^{\rm EX}(\vR) \hspace*{-0.15cm} &=& \hspace*{-0.15cm} - \!\! \sum_{\alpha=p,n} \!
\int \!\! \rho_{\alpha}(\vrr,\vrr-\vs)
            g^{\rm EX}_{p\alpha}(s;\rho_{\alpha}) 
            j_0(K(R) s)
            d \vrr ,~~~~
            \label{U-EX}
\eea
where $V_{\rm CE}$ ($W_{\rm CE}$) is the real (imaginary) part of $U_{\rm CE}$ 
and $\vs=\vrr - \vR $, and $\vrr$ is the coordinate of an interacting nucleon 
from the center-of-mass (c.m.) of target. 
The local momentum $\hbar K(R) \equiv \sqrt{2\mu_{\!R}^{} (E - U_{\rm
CE}-V_{\rm Coul})}$ present in 
Eq.~\eqref{U-EX} is obtained self-consistently, where $\mu_{\! R}^{}$ is 
the reduced mass of the proton+target system. 

The direct and exchange parts, $g^{\rm DR}_{p\a}$ and 
$g^{\rm EX}_{p\a}$, depend on  the local density
\bea
 \rho_{\a}=
  \rho^{\nu}_{\rm P}(\vrr_{\rm T}+\vs/2)
 \label{local-density approximation-SFM}
\eea
at the midpoint of the interacting nucleon pair. 

The direct part $g^{\rm DR}_{p\alpha}$ and the exchange part $g^{\rm EX}_{p\alpha}$ of the $g$-matrix interaction are assumed to be a function 
of the local density $\rho_{\alpha}=\rho_{\alpha}(\vrr-\vs/2)$
at the midpoint of the interacting nucleon pair.
The direct and exchange parts are described as
\bea
&g_{pp}^{\rm DR,EX}(s;\rho_p) = \displaystyle\frac{1}{4} \left( \pm g^{01} + 3
 g^{11}\right) ,\\
 &g_{pn}^{\rm DR,EX} (s;\rho_n) = \displaystyle\frac{1}{8} \left( g^{00} \pm
 g^{01} \pm 3 g^{10} + 3 g^{11}\right) ,
\eea
in terms of the spin-isospin component $g^{ST}$ of the $g$-matrix interaction. 


Similar derivation is possible for the spin-orbit part 
$U_{\rm LS}$~\cite{CEG07}, 
\bea
U_{\rm LS} \equiv V_{\rm LS}+iW_{\rm LS}
=U_{\rm LS}^{\rm DR}+U_{\rm LS}^{\rm EX}
\eea
with 
\bea
&& U_{\rm LS}^{\rm DR} (\vR) = \! - \frac{1}{4 R^2} \!\!\! \sum_{\alpha=p,n} \! \int \!\! \vR \! \cdot
\! \vs \rho_{\alpha} (\vrr) g^{\rm DR}_{{\rm LS},p\alpha}(s;\rho_{\alpha})d\vrr ,\\
&& U_{{\rm LS}}^{\rm EX} (\vR) = -\pi \sum_{\alpha=p,n}\int ds ~s^3 \left[
\frac{2j_0(K(R)s)}{R}\rho_1^{(\alpha)}(R,s) \right. \nonumber \\
&&~~~~~~~~~~~~~~~~~~~~~~~~~~~~~\left. + \frac{j_1(K(R)s)}{2k} \delta_0^{(\alpha)}(R,s) \right] ,
\eea 
where $V_{\rm LS}$ ($W_{\rm LS}$) is the real (imaginary) part 
of $U_{\rm LS}$; see Appendix \ref{Definitions} for the definition of 
$\rho_1^{(\alpha)}(R,s)$ and $\delta_0^{(\alpha)}(R,s)$.
As a $g$-matrix, we take the Kyushu  $g$-matrix resented in Ref.~\cite{Toyokawa:2017pdd}.

The matter radius $r_{\rm m}$ is defined 
with the ground-state wave function $\Phi_{\rm g.s.}$ of T  as 
\begin{align}
\bm r_{\rm m}^2 
&\equiv
 \langle \Phi_{\rm g.s.}|\sum_{i} \bm r_i^2|\Phi_{\rm g.s.}\rangle,
\notag
\\
& =\int d\bm r_{\rm m'}  \langle \Phi_{\rm g.s.}|\sum_{i} \bm r_i^2
 \delta(\bm r_i - \bm r_{\rm m'})|\Phi_{\rm g.s.}\rangle
\notag
\\
& = \int d\bm r_{\rm m'}  \bm r_{\rm m'}^2~\rho( \bm r_{\rm m'}) \label{eq:amddens1}
\end{align}
for the one-body matter density 
\bea
\rho( \bm r_{\rm m})= \langle \Phi_{\rm g.s.}|\sum_{i}^{A} 
 \delta(\bm r_i - \bm r_{\rm m})|\Phi_{\rm g.s.}\rangle 
\eea 
Proton one-body density is 
\bea
\rho( \bm r_{\rm p})= \langle \Phi_{\rm g.s.}|\sum_{j}^{Z} 
 \delta(\bm r_j - \bm r_{\rm p})|\Phi_{\rm g.s.}\rangle,  
\eea
where $j$ is the j-th proton.  
Neutron one-body density is 
\bea
\rho( \bm r_{\rm n})= \langle \Phi_{\rm g.s.}|\sum_{j'}^{N} 
 \delta(\bm r_{j'} - \bm r_{\rm n})|\Phi_{\rm g.s.}\rangle,  
\eea
where $j'$ is the j'-th neutron.

The $S$-matrices are obtained by solving Eq.~\eqref{schrodinger}.  
This point is explained simply by switching off the spin-orbit potential, 
since the spin-orbit potential little affects $\sigma_{\rm R}$.
The $\sigma_{\rm R}$ are determined from the $S$-matrices as 
\bea
\sigma_{\rm R}=\frac{\pi}{k^2} \sum_{L} (2L+1)(1-|S_L|^2),
\eea
where $k$ is a wave number and $L$ is an angular momentum with respect to $\bm R$.

The $U_{\rm CE}$ depends on $\rho( \bm r_{\rm p})$ and $\rho( \bm r_{\rm n})$, as shown in Eq.~\eqref{U-DR} an Eq.~\eqref{U-EX}.
The $S$-matrices depends on the $U_{\rm CE}$.  
The $\sigma_{\rm R}$ depends on the $S$-matrices. 

There is an empirical relation $\sigma_{\rm R}=c~r_{\rm m}^2$ 
between  $r_{\rm m}$ and $\sigma_{\rm R}$, 
Note that  a constant~$c$ is obtained from the calculated 
 $\sigma_{\rm R}$ and $r_{\rm m}$.

\subsection{Double-folding model}

Nucleus-nucleus (PT) elastic scattering can be 
described by the one-body Schr\"odinger equation,
\bea
(T_R + U_{\rm CE} ({\bm R})+V_{\rm Coul}-E){\Psi}^{(+)}=0 
\label{schrodinger-PT}
\eea
with the central; folded potential $U(_{\rm CE}{\bm R})$ shown blow.
Solving Eq.~\eqref{schrodinger-PT}, we can get the S-matrices. 

The folded potential $U_{\rm CE}(\vR)$ 
is composed of the direct and exchange part, $U^{\rm DR}+U^{\rm EX}$, 
defined by
\bea
\label{eq:UD}
U^{\rm DR}(\vR) \hspace*{-0.15cm} &=& \hspace*{-0.15cm} 
\sum_{\mu,\nu}\int \rho^{\mu}_{\rm P}(\vrr_{\rm P}) 
            \rho^{\nu}_{\rm T}(\vrr_{\rm T})
            g^{\rm DR}_{\mu\nu}(s;\rho_{\mu\nu}) d \vrr_{\rm P} d \vrr_{\rm T}, 
            \label{U-DR-DFM}
            \\
U^{\rm EX}(\vR) \hspace*{-0.15cm} &=& \hspace*{-0.15cm}\sum_{\mu,\nu} 
\int \rho^{\mu}_{\rm P}(\vrr_{\rm P},\vrr_{\rm P}-\vs)
\rho^{\nu}_{\rm T}(\vrr_{\rm T},\vrr_{\rm T}+\vs) \nonumber \\
            &&~~\hspace*{-0.5cm}\times g^{\rm EX}_{\mu\nu}(s;\rho_{\mu\nu}) \exp{[-i\vK(\vR) \cdot \vs/M_A]}
            d \vrr_{\rm P} d \vrr_{\rm T},~~~~
            \label{U-EX-DFM}
\eea
where $\vs=\vrr_{\rm P}-\vrr_{\rm T}+\vR$ for the coordinate $\vR$ between P and T and $M_A=A A_{\rm T}/(A +A_{\rm T})$
for the mass number $A$ ($A_{\rm T}$) of P (T). 
The coordinate $\vrr_{\rm P}$ ($\vrr_{\rm T}$) stands for the location 
of an interacting nucleon 
from the center-of-mass of the projectile (target). 
The projectile (target) densities $\rho^{\mu}_{\rm P}(\vrr_{\rm P})$ $(\rho^{\mu}_{\rm T}(\vrr_{\rm T})$) has the $z$ component $\mu$ ($\nu$) of the isospin. 
The local momentum $\hbar K(R) \equiv \sqrt{2\mu_{\!R}^{} (E - U_{\rm
CE}-V_{\rm Coul})}$ present in 
Eq.~\eqref{U-EX-DFM} is obtained self-consistently, where 
$\mu_{\! R}^{}$ is the reduced mass of the PT system. 

The direct and exchange parts, $g^{\rm DR}_{\mu\nu}$ and 
$g^{\rm EX}_{\mu\nu}$, depend on  the local density
\bea
 \rho_{\mu\nu}=\sigma^{\mu} 
 \left[
 \rho^{\nu}_{\rm T}(\vrr_{\rm T}+\vs/2)+\rho^{\nu}_{\rm P}(\vrr_{\rm P}-\vs/2)
 \right]
\label{local-density approximation}
\eea
at the midpoint of the interacting nucleon pair, where $\sigma^{\mu}$ is the Pauli matrix of a nucleon in P.

The direct and exchange parts are described by
\begin{align}
&\hspace*{0.5cm} g_{\mu\nu}^{\rm DR}(s;\rho_{\mu\nu}) \nonumber \\ 
&=
\begin{cases}
\displaystyle{\frac{1}{4} \sum_S} \hat{S}^2 g_{\mu\nu}^{S1}
 (s;\rho_{\mu\nu}) \hspace*{0.42cm} ; \hspace*{0.2cm} 
 {\rm for} \hspace*{0.1cm} \mu+\nu = \pm 1 
 \vspace*{0.2cm}\\
\displaystyle{\frac{1}{8} \sum_{S,T}} 
\hat{S}^2 g_{\mu\nu}^{ST}(s;\rho_{\mu\nu}), 
\hspace*{0.2cm} ; \hspace*{0.2cm} 
{\rm for} \hspace*{0.1cm} \mu+\nu = 0 
\end{cases}
\\
&\hspace*{0.5cm}
g_{\mu\nu}^{\rm EX}(s;\rho_{\mu\nu}) \nonumber \\
&=
\begin{cases}
\displaystyle{\frac{1}{4} \sum_S} (-1)^{S+1} 
\hat{S}^2 g_{\mu\nu}^{S1} (s;\rho_{\mu\nu}) 
\hspace*{0.34cm} ; \hspace*{0.2cm} 
{\rm for} \hspace*{0.1cm} \mu+\nu = \pm 1 \vspace*{0.2cm}\\
\displaystyle{\frac{1}{8} \sum_{S,T}} (-1)^{S+T} 
\hat{S}^2 g_{\mu\nu}^{ST}(s;\rho_{\mu\nu}) 
\hspace*{0.2cm} ; \hspace*{0.2cm}
{\rm for} \hspace*{0.1cm} \mu+\nu = 0 ~~~~~
\end{cases}
\end{align}
where $\hat{S} = {\sqrt {2S+1}}$ and $g_{\mu\nu}^{ST}$ are 
the spin-isospin components of the $g$-matrix interaction.

The Kyushu $g$-matrix~\cite{Toyokawa:2017pdd} is constructed from the chiral interactions with the cutoff 550~MeV.  
In the LF folding model, the chiral $g$-matrix is replaced by the LF $t$ matrix. For a $^{12}$C target, we use the phenomenological density of Ref.~\cite{C12-density} with $r_{\rm m} =2.338$~fm.

The $\sigma_{\rm R}$ are determined from the $S$-matrices as 
\bea
\sigma_{\rm R}=\frac{\pi}{k^2} \sum_{L} (2L+1)(1-|S_L|^2),
\eea
where $k$ is a wave number and $L$ is an angular momentum with respect to $\bm R$. Glauber-model calculations 
regard $\sigma_{\rm R}$ as $\sigma_{\rm I}$, when the data are 
$\sigma_{\rm I({\rm exp})}$. We take the same assumption.

The $U_{\rm CE}$ are obtained from proton and neutron densities of P and T, as shown in Eq.~\eqref{U-DR-DFM} an Eq.~\eqref{U-EX-DFM}.
Solving Eq.~\eqref{schrodinger-PT}, one can get the $S$-matrices. The $S$-matrices yield the $\sigma_{\rm R}$. 

 There is an empirical relation  
$\sigma_{\rm R}=c~[ r_{\rm m}({\rm P})+r_{\rm m}({\rm T}) ]^2$ between $\sigma_{\rm R}$ and $r_{\rm m}({\rm P})$ . 
Note that  a constant~$c$ is obtained from  
 $\sigma_{\rm R}$ and $r_{\rm m}({\rm P})$.

\subsection{Scaling procedure}
\label{Scaling procedure}

The scaled density $\rho_{\rm scaling}(\vrr)$ is obtained 
from the original projectile density $\rho(\vrr)$ as
\bea
\rho_{\rm scaling}(\vrr)=\frac{1}{\a^3}\rho(\vrr/\a) 
\label{eq:scaling}
\eea
with a scaling factor
\bea
\a=\sqrt{ \frac{\langle \vrr^2 \rangle_{\rm scaling}}{\langle \vrr^2 \rangle}} .\eea

In general, we scale 
the D1S-GHFB+AMP or SLy7 proton and neutron densities with Eq.~\eqref{eq:scaling} 
so as to reproduce $\sigma_{\rm R}({\rm exp})$ under  the condition 
that  $r_{\rm p,scaling}=r_{\rm p}({\rm exp})$ of Refs.~\cite{Angeli:2013epw,Brown:2013mga}.
The scaled neutron and proton radii  thus uniquely determined  
are the experimental values.

We assume $A=208$ for $^{nat}$Pb. 
For p+$^{nat}$Pb  scattering, we scale 
the D1S-GHFB+AMP neutron density only 
to the scaled neutron density having $r_{\rm n}({\rm PREX\text{-}II})$
by using Eq.~\eqref{eq:scaling}, since 
$r_{\rm p}({\rm D1S-GHFB+AMP})=r_{\rm p}({\rm PREX\text{-}II})$.

As for scattering of Na and Ar isotopes and $^{24}$Mg on a $^{12}$C target at around 800~MeV/u, we use the LF $t$-matrix folding model.  
In this case, we introduce   a fine-tuning factor $F$ and 
scale the proton and neutron densities 
so as to reproduce $\sigma_{\rm I}({\rm exp})=F \sigma_{\rm R}({\rm LF})$ under the condition that  $r_{\rm p,scaling}=r_{\rm p}({\rm exp})$.

We use the ESP-f of Ref.~\cite{Wakasa:2022ite} 
for $\sigma_{\rm R}$ of p+$^{40}$Ar scattering. 
Now, we explain the ESP-f. The fine-tuning factor $f$ is obtained by averaging  $\sigma_{\rm R}({\rm exp})/\sigma_{\rm R}({\rm th})$ over $E_{\rm lab}$. 
The $f~\sigma_{\rm R}({\rm th})$ is scaled so as to reproduce the data, where the $\sigma_{\rm R}({\rm th})$  is calculated with 
the Kyushu $g$-matrix folding model with reliable proton and neutron densities.

The $g$ matrices include in-medium effects, but the  $t$ matrices do not. In-medium effects are negligible when $E_{\rm lab} \gsim 550$~MeV/u for the $\sigma_{\rm R}$ of 
$^{12}$C+$^{12}$C scattering~\cite{PhysRevC.110.064610}.
The scaling procedure changes the folding potential, but 
the scaling factor $\a$ is close to 1. Therefore,
the Kyushu $g$-matrix folding model has almost  in-medium effects at $\a_{\rm n}=1$; for example $\a_{\rm n}=0.986$ for $^{nat}$Pb.

\section{Results}
\label{Results} 

\subsection{Determination of $r_{\rm skin}^{nat-Pb}$ from the 
reaction cross sections}

\begin{figure}[H]
\begin{center}
 \includegraphics[width=0.5\textwidth,clip]{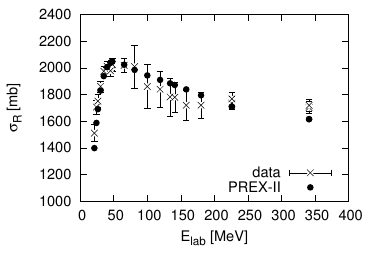}
 \caption{ 
 $E_{\rm lab}$ dependence of $\sigma_{\rm R}$  for p+$^{nat}$Pb scattering. 
Closed circles denote $\sigma_{\rm R}({\rm PREX\text{-}II})$. 
The data are taken from Ref.~\cite{Dietrich:2002swm,RENBERG197281,Cassels1954,Kirkby,
GOLOSKIE1962474,PhysRev.117.1334,GOODING1959241}. 
Note that the cross sections calculated with the scaled densities are not shown, since the cross sections with the scaled densities reproduce the data.
   }
 \label{Fig-RXsec-p+Pb}
 \end{center}
\end{figure}

 Figure~\ref{Fig-RXsec-p+Pb} shows  $E_{\rm lab}$ dependence 
of $\sigma_{\rm R}$ for p+$^{nat}$Pb scattering  in $E_{\rm lab}=21.1 \sim 341$~MeV.  
We assume $A=208$ for $^{nat}$Pb.
The  $\sigma_{\rm R}({\rm PREX\text{-}II})$ are calculated 
with the Kyush $g$-matrix folding model with the D1S-GHFB+AMP neutron density scaled to 
$r_{\rm n}^{208}({\rm PREX\text{-}II})$, where 
$r_{\rm p}^{208}({\rm D1S})=r_{\rm p}^{208}({\rm PREX\text{-}II})$.
The $\sigma_{\rm R}({\rm PREX\text{-}II})$ almost reproduce the central value of the data~\cite{Dietrich:2002swm,RENBERG197281,Cassels1954,Kirkby,
GOLOSKIE1962474,PhysRev.117.1334,GOODING1959241} for each $E_{\rm lab}$.

Scaling the PREX\text{-}II  neutron density 
for $^{nat}$Pb to the data~\cite{Dietrich:2002swm,RENBERG197281,Cassels1954,Kirkby,
GOLOSKIE1962474,PhysRev.117.1334,GOODING1959241} in $E_{\rm lab}=21.1 \sim 341$~MeV. 
we can obtain $r_{\rm skin}^{nat-Pb}=0.271 \pm 0.008$~fm. 
The value agree with $r_{\rm skin}^{208Pb}({\rm PREX\text{-}II})$. 
The values of 
$r_{\rm skin}$ and $r_{\rm m}(\sigma_{\rm R})$ are 
 shown in  Table~\ref{TW values-Pb}. 
The detailed procedure of the scaling  will be explained in Sec.~\ref{Sec;Tb}.

\begin{table}[hbtp]
\caption
{Values  on $r_{\rm m}(\sigma_{\rm R})$, $r_{\rm p}$,$r_{\rm n}$, $r_{\rm skin}$  for $^{nat}$Pb based on proton scattering, where we assume $A=208$.
The radii are shown in units of fm.  
 }
 \begin{tabular}{|c|c|c|c|c|c|c|c|}
 \hline
$A$ & $r_{\rm skin}$ & error & $r_{\rm m}(\sigma_{\rm R})$ & error & $r_{\rm n}$ & error & $r_{\rm p}$  \cr
 \hline
 \hline
208 & 0.271  & 0.008  & 5.610 &	0.005  & 5.715   & 0.008  & 5.444 \cr   
 \hline
 \end{tabular}
 \label{TW values-Pb}
\end{table}

In Refs.~\cite{Tagami:2020bee,Wakasa:2022ite}, we extracted   
$r_{\rm skin}^{208}=0.278 \pm 0.035$~fm from  
$\sigma_{\rm R}$ on p+$^{208}$Pb in $20 \lsim E_{\rm lab} \lsim 180$~MeV. Our present value $r_{\rm skin}=0.271 \pm 0.008$~fm for $^{nat}$Pb agrees with our previous value 
$r_{\rm skin}^{208}=0.278 \pm 0.035$~fm for  $^{208}$Pb.

\subsection{Determination of $r_{\rm skin}$ and $r_{\rm m}$ for $^{159}$Tb}
\label{Sec;Tb}

We extract  $r_{\rm skin}$ and $r_{\rm m}(\sigma_{\rm R})$ from  the data $\sigma_{\rm R}$~\cite{ABEGG1979109}  in $20  \leq E_{\rm lab} \leq 47.5$~MeV, using the Kyushu $g$-matrix folding model with the  D1S-GHFB densities.  $^{159}$Tb is a deformed nucleus;
the deformation parameter $\b_2$ 
calculated with D1S-GHFB is 0.309.

\begin{figure}[H]
\begin{center}
 \includegraphics[width=0.5\textwidth,clip]{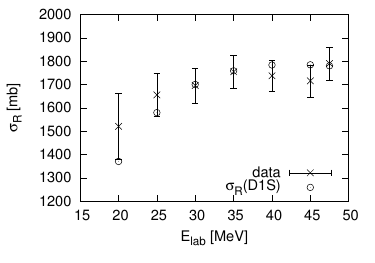}
 \caption{$E_{\rm lab}$ dependence of $\sigma_{\rm R}$ 
 for p+$^{159}$Tb scattering.  
 Open circles stand for the $\sigma_{\rm R}$ calculated with the Kyushu $g$-matrix folding  model with the D1S-GHFB (original) densities. 
 The data are taken from Refs.~\cite{ABEGG1979109}. 
Note that the cross sections calculated with the scaled densities are not shown, since the cross sections with the scaled densities reproduce the data.
    }
 \label{Fig-Xsec-p+Tb}
\end{center}
\end{figure}

Figure \ref{Fig-Xsec-p+Tb} shows our  $\sigma_{\rm R}$  based on the Kyushu $g$-matrix folding model with the  D1S-GHFB densities and  the data~\cite{ABEGG1979109}. 
The  Kyushu $g$-matrix folding  model  (open circles) almost reproduces the central value of the data for each $E_{\rm lab}$.

Taking the scaling procedure of Sec.~\ref{Scaling procedure} for the D1S-GHFB proton and neutron densities, i.e., 
 for each $E_{\rm lab}$, we scale $r_{\rm n}(E_{\rm lab})$ so as to  $\sigma_{\rm R}({\rm scaled})=\sigma_{\rm R}({\rm exp})$
under $r_{\rm p,scaled}=r_{\rm p}({\rm exp})$ of electron scattering~\cite{Angeli:2013epw}. 
We next take a weighted average of  $r_{\rm n}(E_{\rm lab})$ over $E_{\rm lab}$. 
The resultant value is  $r_{\rm n}(\sigma_{\rm R})$. 
We finally obtain $r_{\rm skin}=r_{\rm n}(\sigma_{\rm R})-r_{\rm p}({\rm exp})$ and $r_{\rm m}(\sigma_{\rm R})$; note that 
$A r_{\rm m}(\sigma_{\rm R})^2=N r_{\rm n}(\sigma_{\rm R})^2+Z r_{\rm p}({\rm exp})^2$ . 
Our results based on the scaled proton and neutron densities are  shown in  Table~\ref{TW values-Tb}.

\begin{table}[hbtp]
\caption
{Values  of  $r_{\rm m}(\sigma_{\rm R})$, $r_{\rm p}$,$r_{\rm n}$, $r_{\rm skin}$  for $^{159}$Tb.
The radii are shown in units of fm.  
 }
 \begin{tabular}{|c|c|c|c|c|c|c|c|}
 \hline
$A$ & $r_{\rm skin}$ & error & $r_{\rm m}(\sigma_{\rm R})$ & error & $r_{\rm n}$ & error & $r_{\rm p}$  \cr
 \hline
 \hline
159 & 0.273  & 0.073  & 5.160 &	0.044  & 5.270   & 0.073  & 4.997 \cr   
 \hline
 \end{tabular}
 \label{TW values-Tb}
\end{table}

\subsection{Mg isotopes}

We first explain ``D1S-GHFB+AMP'' in the spherical limit''.
The D1S Hamiltonian is diagonalized in the space spanned 
by the spherical Harmonic oscillator bases. Using 
the wave function thus obtained, we calculate the matter radius.

\begin{figure}[H]
\begin{center}
 \includegraphics[width=0.5\textwidth,clip]{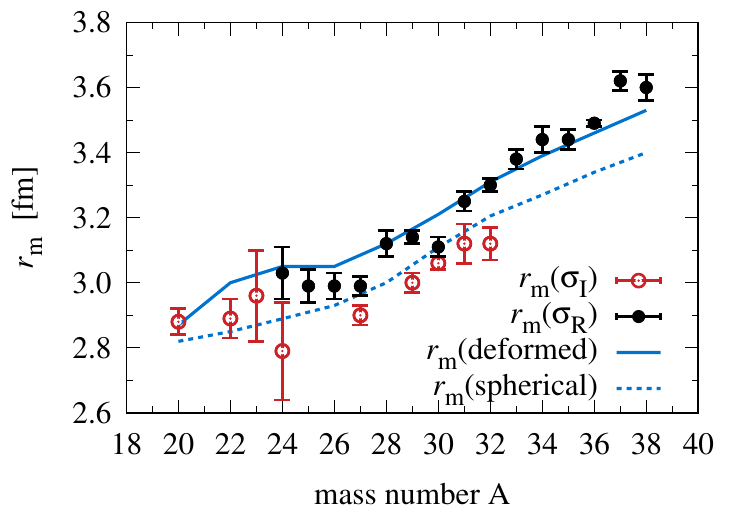}
 \caption{$A$ dependence of  $r_{\rm m}$ for Mg  isotopes. 
 Open circles  with error-bar denote the experimental $r_{\rm m}(\sigma_{\rm I})$ of Ref.~\cite{Suzuki:1998ixv} 
 and  closer circles with error-bar correspond to  
 the experimental $r_{\rm m}(\sigma_{\rm R})$ of Ref.~\cite{Watanabe:2014zea}. 
 For even-even Mg isotopes, the solid line stands for the theoretical $r_{\rm m}({\rm deformed}) $ calculated with D1S-GHFB+AMP, 
 whereas the dashed line corresponds to the theoretical $r_{\rm m}({\rm spherical}) $ calculated with  D1S-GHFB in the spherical limit.
     }
 \label{Fig-Rm-Mg}
\end{center}
\end{figure}

In Fig.~\ref{Fig-Rm-Mg}, the experimental matter radii $r_{\rm m}(\sigma_{\rm I})$ 
of Ref~~\cite{Suzuki:1998ixv}
are compared with the experimental matter radii
$r_{\rm m}(\sigma_{\rm R})$ of Ref.~\cite{Watanabe:2014zea} for $^{24\text{--}38}$Mg. 
The theoretical deformed matter radii 
$r_{\rm m}({\rm deformed}) $ (the solid line) 
calculated with D1S-GHFB+AMP are closer to 
the experimental $r_{\rm m}(\sigma_{\rm R})$, 
whereas the theoretical $r_{\rm m}({\rm spherical}) $ (the dashed line) calculated with  D1S-GHFB in the spherical limit are near the central value of  the experimental $r_{\rm m}(\sigma_{\rm I})$. 
 The experimental $r_{\rm m}(\sigma_{\rm I})$ of Ref~~\cite{Suzuki:1998ixv}
yield matter radii in the spherical limit. 
For $^{24\text{--}38}$Mg, the experimental $r_{\rm m}(\sigma_{\rm R})$ of Ref.~\cite{Watanabe:2014zea} include effects of deformation.

For $^{24}$Mg, the $r_{\rm m}(\sigma_{\rm I})=2.79 \pm 0.15$~fm of Ref.~\cite{Suzuki:1998ixv} and the  $r_{\rm m}(\sigma_{\rm R})=3.03	\pm 0.08$~fm of Ref.~\cite{Watanabe:2014zea}
were extracted from  $^{24}$Mg+$^{12}$C scattering,
Using $\sigma_{\rm I,R}({\rm exp})=c [r_{\rm m}(^{24}{\rm Mg})+r_{\rm m}(^{12}{\rm C})]^2$ for a constant $c$, we can obtain
$(\sigma_{\rm R}-\sigma_{\rm I})/\sigma_{\rm R}=0.087$.
For a $^{12}$C target, we use the phenomenological density of Ref.~\cite{C12-density} with $r_{\rm m}(^{12}{\rm C}) =2.338$~fm.

As a further analysis, we consider whether 
$(\sigma_{\rm R}-\sigma_{\rm I})/\sigma_{\rm R}=0.087$ for $^{24}$Mg+$^{12}$C scattering
 stems from effects of deformation only. 
In Eq.~\eqref{Eq:inela}, the $\sigma_{\rm inela}$ are the total inelastic cross sections. 
When a projectile (P) has large deformation, the $\sigma_{\rm inela}$ are not negligible compared with $\sigma_{\rm I}$. 
There is an empirical formula between $\sigma_{\rm R}$ and $r_{\rm m}({\rm P})$:
\bea
\sigma_{\rm R}=c~[ r_{\rm m}({\rm P})+r_{\rm m}({\rm T}) ]^2 .
\eea
There is an approximation equation having $\b_2$ for 
$r_{\rm m}({\rm P})$:
\bea
r_{\rm m}^{2}({\rm P})= r_{\rm m,0}^{2}({\rm P}) \lfloor 1+ \frac{5\b_2^2}{4\pi} \rfloor =r_{\rm m,0}^{2}({\rm P}) \lfloor 1+ x \rfloor ,
\label{eq:m00}
\eea
where $r_{\rm m,0}$ is the matter radius in the spherical limit 
and $r_{\rm m,0}=$2.929~fm for $\b_2=0.48$ calculated with 
D1S-GHFB+AMP.  
Using the two equations, we can obtain 
\bea
\sigma_{\rm R}&=&c~[ r_{\rm m}({\rm P})+r_{\rm m}({\rm T})
 ]^2 ,
 \\
\sigma_{\rm I}&=&c~[ r_{\rm m,0}({\rm P})+r_{\rm m}({\rm T})
 ]^2 , 
\\
\sigma_{\rm inel}&=&2c x [ (1+x) r_{\rm m,0}({\rm P})^2 + 
r_{\rm m,0}({\rm P}) r_{\rm m}({\rm T}) ] .
\label{eq:inela-only}
\eea
The $\sigma_{\rm inel}$ tend to zero in the limit of  
$\b_2=0$ ($x=0$). The $\sigma_{\rm I}$
yield an experimental matter radius in the spherical limit.

In Fig.~\ref{Fig-Rm-Mg}, the theoretical deformed matter radii 
$r_{\rm m}({\rm deformed}) $ (the solid line) calculated with D1S-GHFB+AMP are close to 
the experimental $r_{\rm m}(\sigma_{\rm R})$, 
but the theoretical $r_{\rm m}({\rm spherical})$ (the dashed line) calculated with  D1S-GHFB in the spherical limit overshoots the central value of  the experimental $r_{\rm m}(\sigma_{\rm I})=2.79 \pm 0.15$~fm of Ref.~\cite{Suzuki:1998ixv}. 
Eventually, the value of $(\sigma_{\rm R}-\sigma_{\rm I})/\sigma_{\rm R}$ is 0.037, when we consider effects of deformation only. 

The $(\sigma_{\rm R}-\sigma_{\rm I})/\sigma_{\rm R}$ is 0.037 is 
smaller than the $(\sigma_{\rm R}-\sigma_{\rm I})/\sigma_{\rm R}=0.087$ mentioned above.
The difference comes from the fact that the $r_{\rm m}(\sigma_{\rm I})=2.79 \pm 0.15$~fm of Ref.~\cite{Suzuki:1998ixv} is too 
small compared with the $r_{\rm m,0}({\rm P})=2.929$~fm obtained by Eq.~\eqref{eq:m00}. 
In Fig.~\ref{Fig-Rm-Mg}, in addition,  the 
$r_{\rm m}(\sigma_{\rm I})$ for $^{23}$Mg is larger than 
that for $^{24}$Mg. 
These facts indicate  that 
the central value of $r_{\rm m}(\sigma_{\rm I})=2.79 \pm 0.15$~fm for $^{24}$Mg is  too small.  
The $r_{\rm m}(\sigma_{\rm I})=2.79 \pm 0.15$~fm for $^{24}$Mg 
 may come from a fluctuation.  
In any case, 
$\sigma_{\rm I}$ experiments are expected in order to obtain $r_{\rm m}(\sigma_{\rm I})$. 

It should be noted that D1S-GHFB+AMP in the spherical limit yields $r_{\rm m}({\rm spherical})=2.925$~fm. The $r_{\rm m,0}({\rm P})=2.929$~fm obtained by 
Eq.~\eqref{eq:m00} agrees with $r_{\rm m}({\rm spherical})=2.925$~fm. This indicates that Eq.~\eqref{eq:m00} is good.

The $r_{\rm skin}$values  of Table~\ref{TW-Mg} are calculated from the $r_{\rm m}(\sigma_{\rm R})$ of Ref.~\cite{Watanabe:2014zea} 
with the $r_{\rm p}({\rm exp})$ determined from the charge radii~\cite{Angeli:2013epw}.

\begin{table}[hbtp]
\caption
{Our results of  $r_{\rm m}$,  $r_{\rm n}$, $r_{\rm skin}$, $r_{\rm p}$ 
for  Mg isotopes.  
The $r_{\rm p}({\rm exp})$ are determined from the charge radii~\cite{Angeli:2013epw}. 
The $r_{\rm m}(\sigma_{\rm R})$ are taken from Ref.~\cite{Watanabe:2014zea}. 
The radii are shown in units of fm. 
}
 \begin{tabular}{|c|c|c|c|c|c|c|c|}
 \hline
$A$ & $r_{\rm skin}$ & error & $r_{\rm m}(\sigma_{\rm R})$ & error & $r_{\rm n}({\rm exp})$ & error & $r_{\rm p}({\rm exp})$  \cr
 \hline
 \hline
24 & 0.17 & 0.15   & 3.03 & 0.08  & 3.11    & 0.15  & 2.943  \cr
25 & 0.14 & 0.09   & 2.99  & 0.05  & 3.06  & 0.09  & 2.915   \cr
26 & 0.13  & 0.07  & 2.99  & 0.04  & 3.05   & 0.07  & 2.922  \cr 
 \hline
 \end{tabular}
 \label{TW-Mg}
\end{table}

\subsection{Na isotopes}

As for $^{21,23}$Na, SLy7 yields better agreement with 
the $r_{\rm m}(\sigma_{\rm I})$ of Refs.~\cite{Suzuki:1995yc,Ozawa:2001hb} than D1S-GHFB do.  
This indicates that SLy is better than  D1S-GHFB for the scaling procedure. We then use SLy7 for the scaling procedure. 

Following the Glauber-model calculations of 
Ref.~\cite{Ozawa:2002av}, we introduce a fine-tuning factor $F$ in our folding model. 
In this case, the $F$ is determined so as to reproduce  
the central  value of $r_{\rm m}(\sigma_{\rm I})=2.79 \pm 0.15$~fm~\cite{Suzuki:1995yc} for  
$^{24}$Mg+$^{12}$C scattering in which the SLy7 densities are used for $^{24}$Mg. The resulting value is $F=0.955$.

\begin{figure}[H]
\begin{center}
 \includegraphics[width=0.5\textwidth,clip]{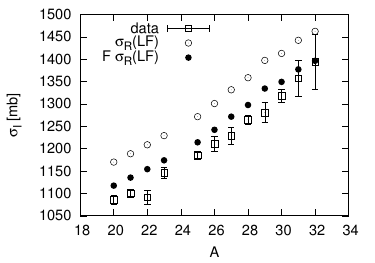}
 \caption{$A$ dependence of $\sigma_{\rm I}$   for Na isotopes. 
 Open circles stand for the results of the  LF $t$-matrix folding  model with the SLy7 densities. 
 Closed circles correspond to the  $F \sigma_{\rm R}({\rm LF})$ 
 with the SLy7 densities with $F=0.955$. 
 The data are taken from Refs.~\cite{Suzuki:1995yc,Ozawa:2001hb}. 
Note that the cross sections calculated with the scaled densities are not shown, since the cross sections with the scaled densities reproduce the data.
   }
 \label{Fig-Xsec-Na}
\end{center}
\end{figure}

Figure \ref{Fig-Xsec-Na} shows $A$ dependence of  $\sigma_{\rm I}$  for Na isotopes. 
The  LF $t$-matrix folding  model with the SLy7 densities overshoots the data~\cite{Suzuki:1995yc}. 
$A$ dependence of  
the data~\cite{Suzuki:1995yc,Ozawa:2001hb} is not natural at $A=22$. 
The $F \sigma_{\rm R}$ calculated with the  LF $t$-matrix folding  model with $F=0.955$ are slightly larger than  the data. 
The $F \sigma_{\rm R}$ are scaled with the scaling procedure of  Sec. \ref{Scaling procedure} and Sec. \ref{Sec;Tb}. 
The resultant values  $r_{\rm m}(\sigma_{\rm I})$ are tabulated in  Table~ \ref{TW values-Na}.

\begin{table}[hbtp]
\caption
{Our values  on  $r_{\rm m}(\sigma_{\rm I})$  for Na isotopes based on 
$\sigma_{\rm I}$. 
The $r_{\rm p}({\rm exp})$ are determined from the charge radii~\cite{Angeli:2013epw}. 
The radii are shown in units of fm.  
 }
 \begin{tabular}{|c|c|c|c|c|c|c|c|}
 \hline
$A$ & $r_{\rm skin}$ & error & $r_{\rm m}(\sigma_{\rm I})$ & error & $r_{\rm n}({\rm exp})$ & error & $r_{\rm p}({\rm exp})$  \cr
 \hline
 \hline
20 & -0.231 & 0.060   & 2.749 & 0.026  & 2.620    & 0.060  & 2.850  \cr
21 & -0.288 & 0.046   & 2.762  & 0.021  & 2.607  & 0.046  & 2.896   \cr
22 & -0.293  & 0.078  & 2.725  & 0.037  & 2.575   & 0.078  & 2.868  \cr 
23 & -0.086  & 0.052 & 2.834   & 0.027  & 2.793   & 0.052  & 2.879  \cr   
25 & 0.039   & 0.035  & 2.887  & 0.020  & 2.904  & 0.035  & 2.865   \cr
26 & 0075  & 0.059  & 2.927    & 0.035  & 2.958   & 0.059  & 2.883  \cr 
27 & 0.070  & 0.064  & 2.949  & 0.039  & 2.977   & 0.064  &2.907  \cr   
28 & 0.133  & 0.034  & 3.017 & 0.021  & 3.069    & 0.034  & 2.936 \cr   
29 & 0.101   & 0.073  & 3.055  & 0.046  & 3.093  & 0.073  & 2.992 \cr
30 & 0.144  & 0.048  & 3.112   & 0.031  & 3.164   & 0.048  & 3.020  \cr 
31 & 0.166  & 0.125  & 3.184   & 0.083  & 3.242  & 0.125  &3.076   \cr   
32 &   &   & 3.255 & 0.121  &    &   &  \cr   
 \hline
 \end{tabular}
 \label{TW values-Na}
\end{table}

\begin{figure}[H]
\begin{center}
 \includegraphics[width=0.45\textwidth,clip]{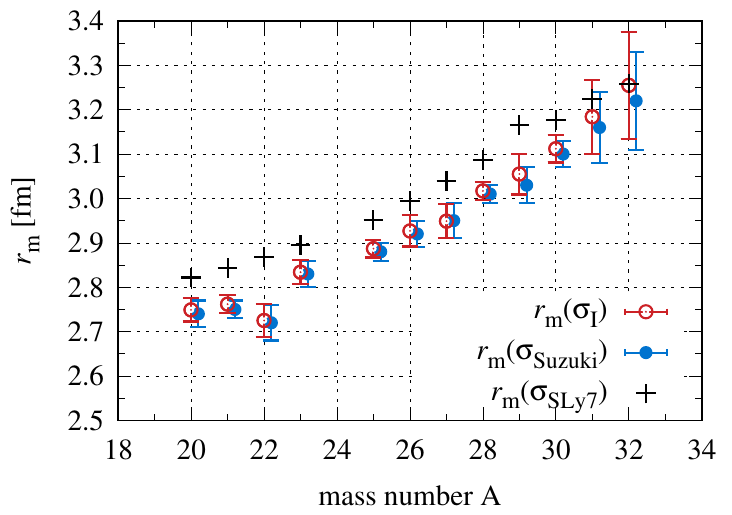}
 \caption{$A$ dependence of our $r_{\rm m}(\sigma_{\rm I})$.~\cite{Suzuki:1995yc,Ozawa:2001hb}, $r_{\rm m}(\sigma_{\rm Suzuki})$  $r_{\rm m}(\sigma_{\rm SLy7})$ for Na isotopes. 
 Open circles with error-bar stand for our $r_{\rm m}(\sigma_{\rm I})$, whereas  closed circles with error-bar correspond to the  $r_{\rm m}(\sigma_{\rm Suzuki})$.~\cite{Suzuki:1995yc,Ozawa:2001hb}. 
The symbol `+' denote $r_{\rm m}(\sigma_{\rm SLy7})$.
    }
 \label{Fig-Rm-Na-SLy7}
\end{center}
\end{figure}

Figure~\ref{Fig-Rm-Na-SLy7} show 
our $r_{\rm m}(\sigma_{\rm I})$, 
the $r_{\rm m}(\sigma_{\rm Suzuki})$ of Refs.~\cite{Suzuki:1995yc,Ozawa:2001hb},  the $r_{\rm m}({\rm SLy7})$ for Na isotopes as a function of $A$. 
Our $r_{\rm m}(\sigma_{\rm I})$ are consistent with the $r_{\rm m}(\sigma_{\rm Suzuki})$. 
The $r_{\rm m}({\rm SLy7})$ are parallel to our $r_{\rm m}(\sigma_{\rm I})$ and 
have natural $A$ dependence, i.e., $r_{\rm m}({\rm SLy7})=1.1263 A^{1/3}-0.3243$ with good accuracy. 
We may say that Sly7 is a good EoS for Na isotopes.

\begin{figure}[H]
\begin{center}
 \includegraphics[width=0.45\textwidth,clip]{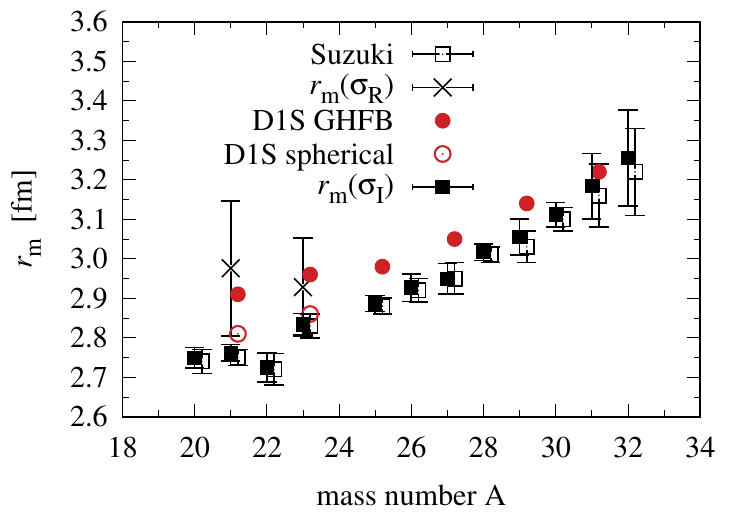}
 \caption{$A$ dependence of  $r_{\rm m}$ for Na  isotopes. 
 Closed circles denote the $r_{\rm m}$ of D1S-GHFB, and open circles correspond 
 to the $r_{\rm m}$ of D1S-GHFB in spherical limit.
Closed squares  with error-bar correspond to our values $r_{\rm m}(\sigma_{\rm I})$, while open squares  with error-bar do to the $r_{\rm m}(\sigma_{\rm I})$ of Refs.~\cite{Suzuki:1998ixv,Ozawa:2001hb}.  
 Crosses with with error-bar show $r_{\rm m}(\sigma_{\rm R})=2.976 \pm 0.171$~fm ($2.928 \pm 0.125$~fm) 
 for $^{21}$Na ($^{23}$Na). 
    }
 \label{Fig-Rm-Na}
\end{center}
\end{figure}

As shown in Fig.~\ref{Fig-Rm-Na}, our values  $r_{\rm m}(\sigma_{\rm I})$ of Table~ \ref{TW values-Na} 
are compared with those of Refs.~\cite{Suzuki:1998ixv,Ozawa:2001hb} for $^{21,23,25,27,29,31}$Na. 
Our values $r_{\rm m}(\sigma_{\rm I})$ agree with those of Refs.~\cite{Suzuki:1998ixv,Ozawa:2001hb}.

Zhang {\it et al.} measured $\sigma_{\rm R}$ for $^{21, 23}$Na+$^{12}$C
scattering at around 30~MeV/u~\cite{ZHANG2002303}. Using  
the Kyushu $g$-matrix with the Sly7 proton and neutron densities and the scaling procedure of Eq.~\eqref{eq:scaling}, we extracted 
$r_{\rm m}(\sigma_{\rm R})=2.976 \pm 0.171$~fm ($2.928 \pm 0.125$~fm) 
from the $\sigma_{\rm R}$ for $^{21}$Na ($^{23}$Na); see Table~\ref{TW values-Na-XSEC=R}. 

\begin{table}[hbtp]
\caption
{Values  on $r_{\rm m}(\sigma_{\rm R})$, $r_{\rm p}$,$r_{\rm n}$, $r_{\rm skin}$  for $^{21,23}$Na.
The radii are shown in units of fm.  
 }
 \begin{tabular}{|c|c|c|c|c|c|c|c|}
 \hline
$A$ & $r_{\rm skin}$ & error & $r_{\rm m}(\sigma_{\rm R})$ & error & $r_{\rm n}$ & error & $r_{\rm p}$  \cr
 \hline
 \hline
21 & 0.283  & 0.334  & 2.976 &	0.171  & 3.121   & 0.334  & 2.838 \cr   
23 & 0.094  & 0.232  & 2.928 &	0.125  & 2.972   & 0.232  & 2.879 \cr   
 \hline
 \end{tabular}
 \label{TW values-Na-XSEC=R}
\end{table}

As shown in Fig.~\ref{Fig-Rm-Na},  for $^{21, 23}$Na, the central values of $r_{\rm m}(\sigma_{\rm R})$ are close to the $r_{\rm m}({\rm D1S})$ (the results of D1S-GHFB), whereas 
the  $r_{\rm m}({\rm D1S-spherical})$ (the results of D1S-GHFB in the spherical limit) are near  our values $r_{\rm m}(\sigma_{\rm I})$ and those of Refs.~\cite{Suzuki:1998ixv,Ozawa:2001hb}.  

For $^{21,23}$Na, the fact that 
the $r_{\rm m}({\rm D1S-spherical})$ are close to our values $r_{\rm m}(\sigma_{\rm I})$ indicates that 
our values $r_{\rm m}(\sigma_{\rm I})$ and the  $r_{\rm m}(\sigma_{\rm I})$ of Refs.~\cite{Suzuki:1998ixv,Ozawa:2001hb} are matter radii in spherical limit. Meanwhile, our $r_{\rm m}(\sigma_{\rm R})$ include 
effects of deformation.

For $^{24\text{--}38}$Mg,  
the $r_{\rm m}({\rm D1S})$ almost agree with the $r_{\rm m}(\sigma_{\rm R})$~\cite{Watanabe:2014zea}  including effects of deformation.   This agreement happens 
for $^{21,23,25,27,29,31}$Na.
For $^{21,23,25,27,29,31}$Na, D1S-GHFB yields 
$\b_2=
0.406, 
0.412, 
0.291, 
0.258,
0.201, 
0.133
$ for $^{21,23,25,27,29,31}$Na, respectively.

\subsection{Ar isotopes}

As for $^{32,34,36,38, 40}$Ar, 
SLy7 yields better agreement with 
the $r_{\rm m}(\sigma_{\rm I})$ of Ref.~\cite{Ozawa:2002av} 
than D1S-GHFB+AMP. 
As the scaling procedure, SLy is better than  D1S-GHFB+AMP. 
This is the reason why we use SLy7.

At $E_{\rm lab}=950$~MeV/u, there is no  $\sigma_{\rm I}$ between Mg isotopes and Ar isotopes. 
Following Ref.~\cite{Ozawa:2002av}, we introduce a fine-tuning factor $F$.   
In this case, the $F$ is determined so as to reproduce  
the central  value of $r_{\rm m}(\sigma_{\rm I})=2.79 \pm 0.15$~fm~\cite{Suzuki:1995yc} for  $^{24}$Mg. The resulting value is $F=0.955$, 
where  
the data $\sigma_{\rm I}({\rm exp})=1136 \pm 72$~mb~\cite{Suzuki:1995yc,Suzuki:1998ixv}  and the the  cross sections calculated with the SLy7 proton and neutron densities is $\sigma_{\rm I}=1252$~mb for 
$^{24}$Mg+$^{12}$C scattering.
The resulting value $F=0.955$ is used also 
for $^{32 \text{--} 40}$Ar+$^{12}$C scattering.

\begin{figure}[H]
\begin{center}
 \includegraphics[width=0.5\textwidth,clip]{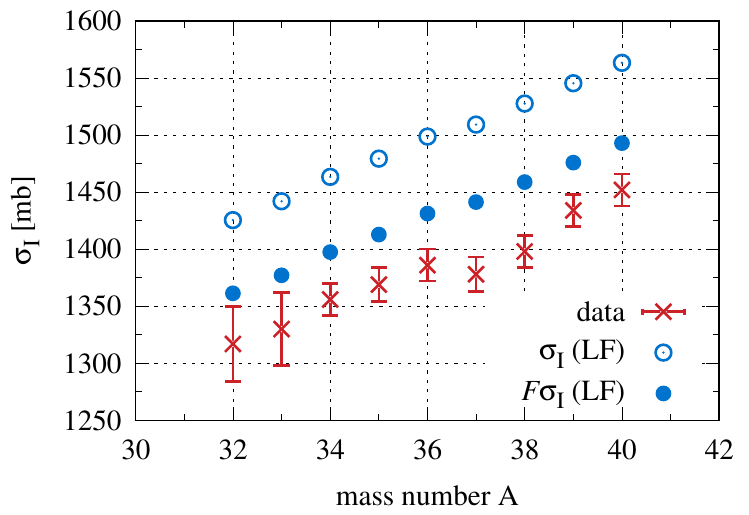}
 \caption{$A$ dependence of $\sigma_{\rm I}$ for Ar isotopes. 
 Open circles stand for the results of the  LF $t$-matrix folding  model with the SLy7 densities. 
 Closed circles correspond to the  $F \sigma_{\rm I}({\rm LF})$ 
 with the SLy7 densities with $F=0.955$. 
 The data is taken from Refs.~\cite{Ozawa:2002av}. 
Note that the cross sections calculated with the scaled densities are not shown, since the cross sections with the scaled densities reproduce the data.
    }
 \label{Fig-Xsec-Ar}
\end{center}
\end{figure}

Figure \ref{Fig-Xsec-Ar} shows $\sigma_{\rm I}$ as a function of $A$.
The  LF $t$-matrix folding  model with the SLy7 densities overestimates
the data~\cite{Ozawa:2002av}. 
The $F \sigma_{\rm I}({\rm LF})$ 
monotonically increase as a function of $A$,
whereas the data~\cite{Ozawa:2002av} have  
a depression in  $^{37,38}$Ar. 
The depression in  $^{37,38}$Ar may be a fluctuation of the data.  
 The $\sigma_{\rm I}$ experiments are expected for Ar isotopes.

The $F~\sigma_{\rm I}$ calculated with the  LF $t$-matrix folding  model with $F$ satisfies  $c F~[ r_{\rm m}({\rm SLy7})+r_{\rm m}({\rm T})]^2$. Except for $A=37,38$,  the $c$ hardly depend on $A$.

The $F \sigma_{\rm R}$ calculated with the  LF $t$-matrix folding  model with $F=0.955$ are slightly larger than  the data. 
The $F \sigma_{\rm R}$ are scaled with the scaling procedure of  Sec. \ref{Scaling procedure} and Sec. \ref{Sec;Tb}. 
The resultant values  $r_{\rm m}(\sigma_{\rm I})$ are tabulated in  Table~ \ref{TW values-Ar}.

\begin{figure}[H]
\begin{center}
 \includegraphics[width=0.5\textwidth,clip]{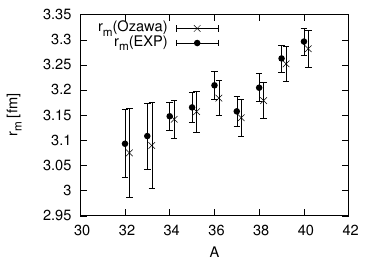}
 \caption{$A$ dependence of  $r_{\rm m}(\sigma_{\rm I})$ for Ar  isotopes. 
 Closed circles with error-bar correspond to our values, while crosses with error-bar does to those of Ref.~\cite{Ozawa:2002av}.  
    }
 \label{Fig-Rm-Ar-0}
\end{center}
\end{figure}

Our values $r_{\rm m}(\sigma_{\rm I})$ are compared with the $r_{\rm m}({\rm Ozawa})$ of Ref.~\cite{Ozawa:2002av}, as shown in Fig.~\ref{Fig-Rm-Ar-0}.
Our results are consistent with those of 
Ref.~\cite{Ozawa:2002av}. 
Both our $r_{\rm m}(\sigma_{\rm I})$ and $r_{\rm m}({\rm Ozawa})$ have a depression in $^{37,38}$Ar.  SLy7 and D1S do not yield the  depression. The reason why our $r_{\rm m}(\sigma_{\rm I})$ and $r_{\rm m}({\rm Ozawa})$ are smaller than the $r_{\rm m}(\sigma_{\rm I})$ of the neighborhood nuclei stems from the fact that the data 
$\sigma_{\rm I}$ of Ref. \cite{Ozawa:2002av} are smaller for $^{37,38}$Ar.

\begin{figure}[H]
\begin{center}
 \includegraphics[width=0.5\textwidth,clip]{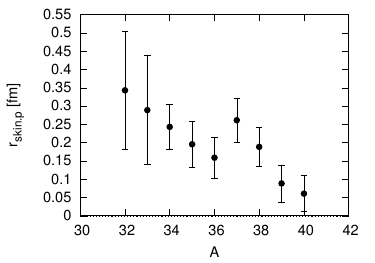}
 \caption{$A$ dependence of  $r_{\rm skin.p}$ for $^{32 \text{--} 40}$Ar.
    }
 \label{Fig-Ar-skins}
\end{center}
\end{figure}

\begin{figure}[H]
\begin{center}
 \includegraphics[width=0.5\textwidth,clip]{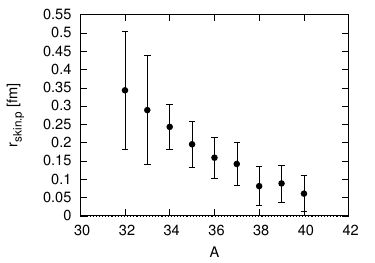}
 \caption{$A$ dependence of  $r_{\rm skin.p}$ for $^{32 \text{--} 40}$Ar 
  in which the upper-bound values are taken for  $A=37, 38$.
    }
 \label{Fig-skins-1}
\end{center}
\end{figure}

Figure  \ref{Fig-Ar-skins} shows  our results on 
proton skin thickness 
$r_{\rm skin.p}$ for $^{32 \text{--} 40}$Ar.
There is a bump at $A=37, 38$. This is related to the fact that 
the central values of $\sigma_{\rm I}({\rm exp})$ are slightly smaller than those of the neighborhood nuclei. 
If we take the upper-bound values
of the data $\sigma_{\rm I}({\rm exp})$ for $A=37, 38$  in  Fig.~\ref{Fig-Xsec-Ar}, we obtain Fig.  \ref{Fig-skins-1} in which the bump almost disappears. 
The bump comes from the depression in  $^{37,38}$Ar shown in  Fig.~\ref{Fig-Xsec-Ar}, The depression in  $^{37,38}$Ar may be a fluctuation of the data.   Our final results  are  the $r_{\rm skin.p}$ of Fig.~ \ref{Fig-Ar-skins}, since it is not easy to avoid 
the fluctuation. The re-measurements are expected.

Our results are tabulated in Table  \ref{TW values-Ar}.
Our values $r_{\rm m}(\sigma_{\rm I})$  are consistent with those of Ref.~\cite{Ozawa:2002av}.

\begin{table}[hbtp]
\caption
{Values  on  $r_{\rm p}$, $r_{\rm m}(\sigma_{\rm I})$, $r_{\rm n}$, $r_{\rm skin,p}$  for Ar isotopes based on $\sigma_{\rm I}$. 
The radii are shown in units of fm.  
 }
 \begin{tabular}{|c|c|c|c|c|c|c|c|}
 \hline
$A$ & $r_{\rm skin}$ & error & $r_{\rm m}(\sigma_{\rm I})$ & error & $r_{\rm n}$ & error & $r_{\rm p}$  \cr
 \hline
 \hline
32 & 0.343 & 0.163  & 3.093 & 0.068  & 2.896   & 0.163  & 3.2387   \cr
33 & 0.289  & 0.149  & 3.109  & 0.065  & 2.948  & 0.149  & 3.2366   \cr
34 & 0.243  & 0.062  & 3.148  & 0.028  & 3.017   & 0.062  & 3.2599  \cr 
35 & 0.195  & 0.064 & 3.166  & 0.030  & 3.064   & 0.064  & 3.2590  \cr   
36 &0.159  & 0.057  & 3.209  & 0.028  & 3.129  & 0.057  & 3.2878  \cr
37 & 0.261  & 0.060  & 3.158  & 0.030  & 3.028  & 0.060  & 3.2891  \cr
38 & 0.188  & 0.054  & 3.205   & 0.028  & 3.114  & 0.054  & 3.3024  \cr 
39 & 0.088  & 0.051  & 3.263  & 0.027  & 3.222  & 0.051  &3.3101  \cr   
40 & 0.061  & 0.049  & 3.296 & 0.027  & 3.269   & 0.049  & 3.3297 \cr   
 \hline
 \end{tabular}
 \label{TW values-Ar}
\end{table}

\subsection{p+$^{40}$Ar scattering}

There is no data on $\sigma_{\rm R}$ for $^{32\text{-}39}$Ar. 
Now, we extract  $r_{\rm m}(\sigma_{\rm R})$ from $\sigma_{\rm R}({\rm exp})$~\cite{CARLSON198557}  for 
p+$^{40}$Ar scattering, using the Kyushu $g$-matrix folding model with the Sly7 proton and neutron densities.

\begin{figure}[H]
\begin{center}
 \includegraphics[width=0.5\textwidth,clip]{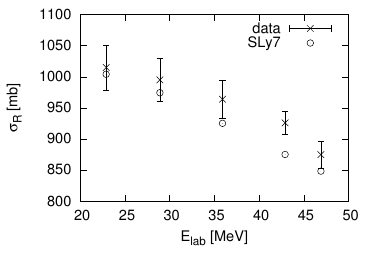}
 \caption{$E_{\rm lab}$ dependence of $\sigma_{\rm R}$ 
 for p+$^{40}$Ar scattering.  
 Open circles stand for results of the Kyushu $g$-matrix folding  model with the SLy7 densities. 
 The data is taken from Refs.~\cite{CARLSON198557}. 
Note that the cross sections calculated with the scaled densities are not shown, since the cross sections with the scaled densities reproduce the data.
   }
 \label{Fig-Xsec-p+Ar}
\end{center}
\end{figure}

Figure \ref{Fig-Xsec-p+Ar} shows our values on $\sigma_{\rm R}$ 
and  the data~\cite{CARLSON198557}. 
The  Kyushu $g$-matrix folding  model with the SLy7 densities almost agrees with the data~\cite{CARLSON198557}. We then introduce     
a fine-tuning factor $f$. 
We use the reliable ESP-f of Ref.~\cite{Wakasa:2022ite} in order to determine $f$.
The fine-tuning factor $f$ is obtained by averaging  $\sigma_{\rm R}({\rm exp})/\sigma_{\rm R}({\rm th})$ over $E_{\rm lab}$. The resulting value is $f=1.0327$. After the scaling procedure based on ESP-f, 
 we obtain $r_{\rm m}(\sigma_{\rm R})$, as shown in Table  \ref{TW values-Ar-p}.
The $r_{\rm m}(\sigma_{\rm R})$  is larger than 
$r_{\rm m}(\sigma_{\rm I})$, as shown in  Table~\ref{TW values-Ar}. 

\begin{table}[hbtp]
\caption
{Values  on $r_{\rm m}(\sigma_{\rm R})$, $r_{\rm p}$,$r_{\rm n}$, $r_{\rm skin}$  for $^{40}$Ar based on proton scattering.
The radii are shown in units of fm.  
 }
 \begin{tabular}{|c|c|c|c|c|c|c|c|}
 \hline
$A$ & $r_{\rm skin}$ & error & $r_{\rm m}(\sigma_{\rm R})$ & error & $r_{\rm n}$ & error & $r_{\rm p}$  \cr
 \hline
 \hline
40 & 0.094  & 0.054  & 3.381 &	0.030  & 3.423   & 0.054  & 3.3297 \cr   
 \hline
 \end{tabular}
 \label{TW values-Ar-p}
\end{table}

\begin{figure}[H]
\begin{center}
 \includegraphics[width=0.5\textwidth,clip]{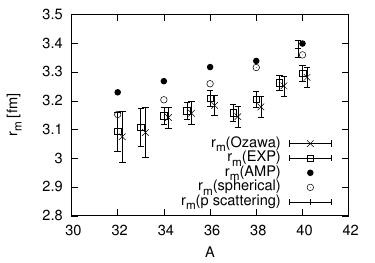}
 \caption{$A$ dependence of  $r_{\rm m}$ for Ar  isotopes. 
 Closed circles denote the $r_{\rm m}$ of . D1S-GHFB+AMP, whereas open circles correspond  to the $r_{\rm m}$ of D1S-GHFB+AMP in spherical limit
Squares with error-bar denote our values $r_{\rm m}(\sigma_{\rm I})$, while crosses with error-bar correspond to the $r_{\rm m}({\rm Ozawa})$ of Ref.~\cite{Ozawa:2002av}.  Symbol ``+'' with an error-bar shows $r_{\rm m}(\sigma_{\rm R;p~scattering})$.
    }
 \label{Fig-Rm-Ar}
\end{center}
\end{figure}

Figure \ref{Fig-Rm-Ar} show $A$ dependence of the $r_{\rm m}(\sigma_{\rm R})$ of Table~\ref{TW values-Ar-p}, the $r_{\rm m}(\sigma_{\rm I})$ of Table~\ref{TW values-Ar}, 
the $r_{\rm m}(\sigma_{\rm I})$ 
of Ref.~\cite{Ozawa:2002av}. 
These experimental values are compared with 
the theoretical matter radii, $r_{\rm m}({\rm AMP})$ 
 ($r_{\rm m}({\rm deformed})$)  and  
$r_{\rm m}({\rm spherical})$,   for $^{32,34,36,38, 40}$Ar, 
where  
the $r_{\rm m}({\rm AMP})$ are the results  calculated with 
D1S-GHFB+AMP, whereas $r_{\rm m}({\rm spherical})$ corresponds to 
D1S-GHFB+AMP in the spherical limit. Note that D1S-GHFB+AMP yields 
$\b_2=
0.244, 
0.225, 
0.219, 
0.149, 
0.203 
$ for $^{32,34,36,38, 40}$Ar, respectively.
Note that the AMP calculations are very difficult for odd nuclei;
see Ref.~\cite{PhysRevC.101.014620} for the difficulty. 
As for $^{38}$Ar, the $\b_2$ is smallest, because of  $N=20$ (closed shell).

For $^{40}$Ar, the $r_{\rm m}({\rm AMP})$ and the $r_{\rm m}({\rm spherical})$ reproduce our $r_{\rm m}(\sigma_{\rm R; p~scattering})$ (Symbol ``+'' with an error-bar) extracted from p+ $^{40}$Ar scattering within the error-bar of $r_{\rm m}(\sigma_{\rm R; p~scattering})$. 
This indicates that  the $r_{\rm m}(\sigma_{\rm R; p~scattering})$ includes effects of deformation, although the effects are small. 
Our value $r_{\rm m}(\sigma_{\rm I})$ for $^{40}$Ar is below 
$r_{\rm m}({\rm spherical})$, indicating that our value $r_{\rm m}(\sigma_{\rm I})$ does not includes effects of deformation.

\section{Summary}
\label{Summar}

We have extracted  the  neutron skin thickness $r_{\rm skin}=0.271 \pm 0.008$~fm for $^{nat}$Pb from $\sigma_{\rm R}$ of p+$^{nat}$Pb scattering in $21 \lsim E_{\rm lab} \lsim 341$~MeV, using the chiral (Kyushu) $g$-matrix folding model, where we assume $A=208$ for $^{nat}$Pb. The chiral (Kyushu) $g$-matrix folding model having 
$r_{\rm p}({\rm PREX\text{-}II})$ and $r_{\rm m}({\rm PREX\text{-}II})$ reproduce the data in $21 \lsim E_{\rm lab} \lsim 341$~MeV. This indicates that the chiral (Kyushu) $g$-matrix folding model is reliable there. 
The  skin value  for $^{nat}$Pb agrees with $r_{\rm skin}^{208}({\rm PREX\text{-}II}) = 0.283\pm 0.071$~fm for  $^{208}$Pb

Meanwhile, in Refs.~\cite{Tagami:2020bee,Wakasa:2022ite}, we extracted   
$r_{\rm skin}^{208Pb}=0.278 \pm 0.035$~fm from  
$\sigma_{\rm R}$ on p+$^{208}$Pb in $20 \lsim E_{\rm lab} \lsim 180$~MeV.
 Our present value $r_{\rm skin}^{nat}=0.271 \pm 0.008$~fm of  $^{nat}$Pb agrees with our previous value $r_{\rm skin}^{208Pb}=0.278 \pm 0.035$~fm of $^{208}$Pb. Prediction A  may be true, where
prediction A is that the skin values  for $^{208,207,206,204}$Pb are almost constant as a function of $A$, as mentioned in Sec.~\ref{Introduction}.

For $^{159}$Tb, our results are $r_{\rm m}=5.160 \pm 
0.044$~fm and $r_{\rm skin}=0.273 \pm 0.073$~fm; note that 
D1S-GHFB shows 
$\b_2=0.309$. 
\bea
r_{\rm m}^{2}= r_{\rm m,0}^{2} [ 1+ \frac{5\b_2^2}{4\pi} ] =r_{\rm m,0}^{2}  
\label{eq:m0}
\eea
where $r_{\rm m,0}$ is the matter radius in the spherical limit. 
Using Eq.~\eqref{eq:m0}, we can get $r_{\rm m,0}=5.065 
 \pm 0.044$~fm and $r_{\rm skin,0}=0.114 	\pm 0.074$~fm in the spherical limit. Effects of deformation are thus large for $r_{\rm skin}$

We have determined matter radii  $r_{\rm m}(\sigma_{\rm R})$ and $r_{\rm skin}(\sigma_{\rm R})$
for $^{40}$Ar and  $r_{\rm m}(\sigma_{\rm I})$ and $r_{\rm skin}(\sigma_{\rm R})$
for ${^{20{\text--}32}}$Na, $^{32{\text--}40}$Ar; see Tables  \ref{TW values-Na},  \ref{TW values-Na-XSEC=R}, \ref{TW values-Ar},  \ref{TW values-Ar-p}. 
We have used the Kyushu  $g$-matrix folding model for low and intermediate energies and the LF $t$-matrix folding model   for high energies. 

As for $^{24}$Mg, our value $r_{\rm m}(\sigma_{\rm R})=3.03	\pm 0.08$~fm~\cite{Watanabe:2014zea} includes effects of deformation, whereas the $r_{\rm m}(\sigma_{\rm I})=2.79 \pm 0.15$~fm~\cite{Ozawa:2001hb}  does not. 

As for even-even Mg isotopes, we have used D1S-GHFB+AMP and 
its spherical limit in order to 
find whether the nucleus we consider includes effects of deformation or not. This analyses show that 
the $r_{\rm m}(\sigma_{\rm I})$ of Ref~~\cite{Suzuki:1998ixv} for 
$^{20, 22, 23, 27, 29\text{--}32}$Mg 
yield matter radii in the spherical limit, whereas the $r_{\rm m}(\sigma_{\rm R})$ for $^{24\text{--}38}$Mg of Ref.~\cite{Watanabe:2014zea} include effects of deformation.

As shown in Fig.~\ref{Fig-Rm-Na-SLy7},  
our   $r_{\rm m}(\sigma_{\rm I})$ are consistent with the $r_{\rm m}(\sigma_{\rm Suzuki})$ of Ref.~\cite{Suzuki:1995yc} for 
$^{20\text{--}23, 25\text{--}32}$Na. 
Our   $r_{\rm m}(\sigma_{\rm I})$ and the $r_{\rm m}(\sigma_{\rm Suzuki})$ are, however, the matter radii from measured $\sigma_{\rm I}$. Our  $r_{\rm m}(\sigma_{\rm I})$ and the $r_{\rm m}(\sigma_{\rm Suzuki})$ should be compared with   matter radii $r_{\rm m}(\sigma_{\rm R})$ determined from  $\sigma_{\rm R}$, when $\sigma_{\rm R}$ are available. 
This is because $\b_2$ calculated with D1S-GHFB are large  for $^{21, 23}$Na; namely $\b_2=0.406, 0.412$ for $^{21, 23}$Na, respectively.

Zhang {\it et al.} measured reaction cross sections 
$\sigma_{\rm R}$ for $^{21, 23}$Na+$^{12}$C
scattering at around 30~MeV/u~\cite{ZHANG2002303}. Using  the Kyushu $g$-matrix with the Sly7 proton and neutron densities and the scaling procedure of Sec. \ref{Scaling procedure} and Sec.\ref{Sec;Tb}., 
we extracted 
$r_{\rm m}(\sigma_{\rm R})=2.976 \pm 0.171$~fm ($2.928 \pm 0.125$~fm) 
from the $\sigma_{\rm R}$ for $^{21}$Na ($^{23}$Na). 

\begin{figure}[H]
\begin{center}
 \includegraphics[width=0.5\textwidth,clip]{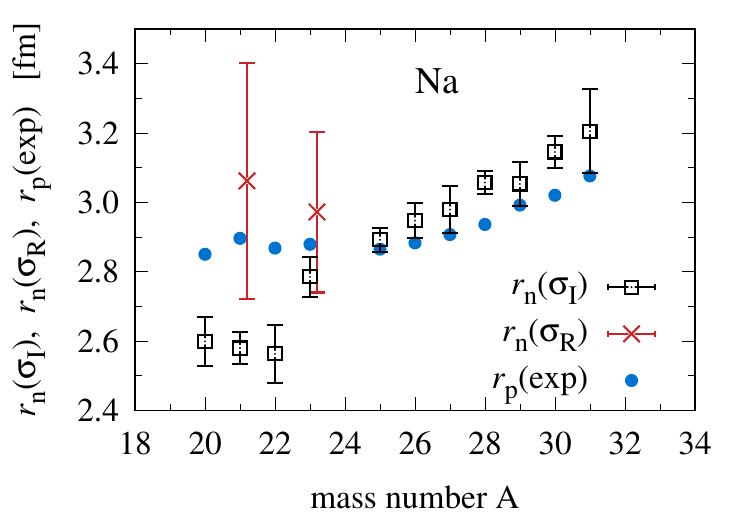}
 \caption{$A$ dependence of  the $r_{\rm n}(\sigma_{\rm I})$, 
 the  $r_{\rm n}(\sigma_{\rm R})$, the $r_{\rm p}({\rm exp})$ for Na  isotopes. 
 Squares with error-bar denote our values for $r_{\rm n}(\sigma_{\rm I})$, while circles with error-bar  correspond to the $r_{\rm p}({\rm exp})$ 
 determined from the charge radii~\cite{Angeli:2013epw}. 
Crosses with with error-bar show our 
 $r_{\rm n}(\sigma_{\rm R})=2.976 \pm 0.171$~fm 
 ($2.928 \pm 0.125$~fm) for $^{21}$Na ($^{23}$Na).
    }
 \label{Fig-Rn-Na}
\end{center}
\end{figure}

For $^{21,23}$Na, our values $r_{\rm n}(\sigma_{\rm R})$ include effects of deformation, but our values $r_{\rm n}(\sigma_{\rm I})$ do not. 
For Na isotopes,  our $r_{\rm n}(\sigma_{\rm R})$ and 
our $r_{\rm n}(\sigma_{\rm I})$, are shown together with $r_{\rm p}({\rm exp})$ in Fig.~\ref{Fig-Rn-Na}. We find that effects of deformation on $r_{\rm n}$ are not negligible. 

As for Ar isotopes, our results  $r_{\rm m}(\sigma_{\rm I})$ are 
consistent with those of Ref.~\cite{Ozawa:2002av} for $^{32{\text--}40}$Ar.
We extract  $r_{\rm m}(\sigma_{\rm R})$ from  the data~\cite{CARLSON198557} 
on $\sigma_{\rm R}({\rm exp})$ for 
p+$^{40}$Ar scattering, using the Kyushu $g$-matrix folding model with the Sly7 densities in which the scaling procedure (ESP-f) of Ref.~\cite{Wakasa:2022ite} is used. For $^{40}$Ar, our value $r_{\rm m}(\sigma_{\rm R})$ is  larger than our values $r_{\rm m}(\sigma_{\rm I})$. 
The former value includes effects of deformation, but the latter value does not.

\section*{Acknowledgements}
We thank Toyokawa for his comments.

\hspace*{1 cm}

\appendix
\section{Definitions of quantities}
\label{Definitions}
The explicit forms of $\delta_0^{(\alpha)}(R,s)$ 
and $\rho_1^{(\alpha)}(R,s)$ are 
\bea
&& \delta_0^{(\alpha)}(R,s)\nonumber \\
&& ~~=\frac{1}{2}\int_{-1}^{+1} d\omega
  \frac{g^{\rm EX}_{{\rm LS},p\alpha}( s;\rho_{\alpha})}{x}
  \nonumber \\
&& ~~~~~\times \left\{\left. \frac{3}{k^{(\alpha)}_F(x) s}
		     j_1(k^{(\alpha)}_F(x) s) \frac{d}{dx}
		     \rho_{\alpha}(x) \right|_{x=\sqrt{R^2+s^2/4+R s
\omega }} \right. \nonumber \\
&& ~~~~~\left. +s\rho_{\alpha}(x) \frac{d}{dx} k^{(\alpha)}_F(x) \right|_{x=\sqrt{R^2+s^2/4+R s
\omega }} \nonumber \\
&& ~~~~~\left.\left. \times \frac{d}{dy}\left[\frac{3}{y}j_1(y)
					    \right] \right|_{y=k^{(\alpha)}_F(x) s} \right\}, 
\eea
and 
\bea
&& \rho_1^{(\alpha)}(R,s) \nonumber \\
&& ~~ =\frac{1}{2}\int_{-1}^{+1}d\omega ~\omega g^{\rm EX}_{{\rm
 LS},p\alpha}(s;\rho_{\alpha})\frac{3}{k^{(\alpha)}_F(x) s} \nonumber \\
&& ~~~~~ \left. \times j_1(k^{(\alpha)}_F(x) s)\rho_{\alpha}(y) \right|_{y=\sqrt{R^2+s^2/4+R s \omega}}, 
\eea
where 
\bea
&& k^{(\alpha)}_F = (3 \pi^2 \rho_{\alpha})^{1/3}, \\
&& g^{\rm DR,EX}_{{\rm LS},pp} (s;\rho_p)= g^{11}_{\rm LS},\\
&& g^{\rm DR,EX}_{{\rm LS},pn} (s;\rho_n)= \frac{1}{2}\left( \pm g^{10}_{\rm LS} + g^{11}_{\rm LS}  \right) .
\eea

\bibliography{Folding-v27}

\end{document}